# Unconventional Meissner screening induced by chiral molecules in a conventional superconductor

**Hen Alpern[1,2], Morten Amundsen[3], Roman Hartmann[4], Nir Sukenik[1], Alfredo Spuri[4], Shira Yochelis[1], Thomas Prokscha[5], Vitaly Gutkin[6], Yonathan Anahory[2], Elke Scheer[4], Jacob Linder[3], Zaher Salman[5], Oded Millo[2], Yossi Paltiel[1], Angelo Di Bernardo[4,*]**

1. Applied Physics Department and the Centre for Nanoscience and Nanotechnology, The Hebrew University of Jerusalem, 91904 Jerusalem, Israel
2. Racah Institute of Physics and the Centre for Nanoscience and Nanotechnology, The Hebrew University of Jerusalem, 91904 Jerusalem, Israel
3. Centre for Quantum Spintronics, Department of Physics, Norwegian University of Science and Technology, NO-7491 Trondheim, Norway.
4. Department of Physics, University of Konstanz, 78457 Konstanz, Germany
5. Laboratory for Muon Spin Spectroscopy, Paul Scherrer Institut, 5232 Villigen PSI, Switzerland
6. The Harvey M. Krueger Family Centre for Nanoscience and Nanotechnology, The Hebrew University of Jerusalem, Jerusalem 91904, Israel

*Corresponding author. Email: angelo.dibernardo@uni-konstanz.de

## Abstract

The coupling of a superconductor to a different material often results in a system with unconventional superconducting properties. A conventional superconductor is a perfect diamagnet expelling magnetic fields out of its volume, a phenomenon known as Meissner effect. Here, we show that the simple adsorption of a monolayer of chiral molecules, which are non-magnetic in solution, onto the surface of a conventional superconductor can markedly change its diamagnetic Meissner response. By measuring the internal magnetic field profile in superconducting Nb thin films under an applied transverse field by low-energy muon spin rotation spectroscopy, we demonstrate that the local field profile inside Nb is considerably modified upon molecular adsorption in a way that also depends on the applied field direction. The modification is not limited to the chiral molecules/Nb interface, but it is long ranged and occurs over a length scale comparable to the superconducting coherence length. Zero-field muon spin spectroscopy measurements in combination with our theoretical analysis show that odd-frequency spin-triplet states induced by the chiral molecules are responsible for the modification of the Meissner response observed inside Nb. These results indicate that a chiral molecules/superconductor system supports odd-frequency spin-triplet pairs due to the molecules acting as a spin active layer and therefore they imply that such system can be used as a simpler alternative to superconductor/ferromagnet or superconductor/topological insulator hybrids for the generation and manipulation of unconventional spin-triplet superconducting states.

# I. INTRODUCTION

The elemental charge unit of a superconductor (S), the Cooper pair of electrons, offers degrees of freedom with respect to its orbital, frequency and spin symmetry. In a conventional superconductor, the Cooper pairs of electrons are condensed into a ground state that is described by a macroscopic wavefunction with a spatially isotropic (*s*-wave) even-frequency and spin-singlet symmetry [1]. One of the hallmarks of such conventional superconducting state is the diamagnetic Meissner screening [2], meaning the expulsion of an applied external field from the interior of a conventional S.

Over the past two decades, it has been demonstrated that *unconventional* superconducting states can be generated from the combination of a conventional S with a different material. One of the most peculiar examples of such combination is that consisting of a S coupled to a ferromagnet (F). Here, the F's exchange field causes a change in the spin symmetry of the Cooper pair wavefunction inside S, which due to the Pauli principle results in the emergence of odd-frequency spin-triplet superconducting states [3-8]. Similarly, it was suggested that a self-assembled monolayer (SAM) of chiral molecules (ChMs) can induce a modification of the superconducting order parameter (OP) of a conventional S onto which it is adsorbed [9-12]. This suggestion is based on the chiral-induced spin selectivity effect that ChMs exhibit due to the preferential transmission of electrons with a certain spin orientation through them [13-14], yielding the ability of ChMs to magnetize Fs [15-17]. Indeed, low-temperature scanning tunneling microscopy [9,10] and transport measurements [11,12] show that zero-energy bound states compatible with odd-frequency spin-triplet superconductivity can emerge in a ChMs/S system. Sparsely adsorbed ChMs also seem to act as magnetic impurities inducing surface Shiba-like states [12, 18]. Although these results suggest that ChMs can modify *locally* the OP of a conventional S [9-12], they do not explain whether the effects observed are just limited to the ChMs/S interface or whether the ChMs can also modify the intrinsic superconducting properties of a S, such as the screening-current distribution in S, even further away from the ChMs/S interface. Moreover, the magnetic spin activity of the ChMs layer coupled to a S, as known to exist at S/F interfaces, has never been demonstrated.

Here, we use low-energy muon spin spectroscopy (LE-μSR) to resolve the impact, in terms of both its depth dependence and magnitude, which a SAM of ChMs has on the intrinsic superconducting screening properties of a conventional S thin film. Our results provide evidence for a global change in the symmetry of the superconducting order parameter and elucidate the physical mechanisms responsible for it. The order parameter modification is evidenced by a radical variation in the screening properties of a S thin film upon ChMs adsorption deep inside the S. We attribute this modification to the emergence of unconventional spin-triplet superconductivity induced by the magnetic spin activity of the ChMs/S interface. The spin activity is demonstrated by LE-μSR measurements in zero-field as well as in traverse-field, where we also show the asymmetry of the unconventional Meissner screening observed in the ChM/S system upon switching of the applied field direction.

## II. UNCONVENTIONAL MEISSNER SCREENING INDUCED BY ChMs

In this study we adsorb a SAM of alpha helix polyalanine ChMs onto the surface of S thin films of Nb (~ 65 nm and ~ 55 nm in thickness). Successful adsorption of ChMs onto the Nb surface was already reported [11], and we also demonstrate it on our Nb samples using scanning Kelvin probe microscopy (see Appendix A, B). After adsorbing ChMs onto a 65-nm-thick Nb film, we use LE-μSR to probe the depth dependence of the local magnetic field profile, $B_{loc}$, inside this sample, and then compare the measured $B_{loc}$ profile with that obtained for a 65-nm-thick Nb film grown in the same deposition run under the same conditions, but without the ChMs. We measure $B_{loc}$ as this is directly correlated to the spatial distribution of the S screening supercurrents, which ultimately depends on the OP inside the S.

LE-μSR allows probing the $B_{loc}$ variation along the muon implantation direction (*z*-axis in Fig. 1) with a sensitivity better than 0.1 Gauss and a depth resolution of a few nanometers [19-24]. Thanks to these unique features, the LE-μSR technique has already been used, for example, to resolve the depth variation of $B_{loc}$ and detect an anomalous Meissner response due to the formation of unconventional superconducting states in a S coupled to a F [8,22-23] or to a topological insulator [24].

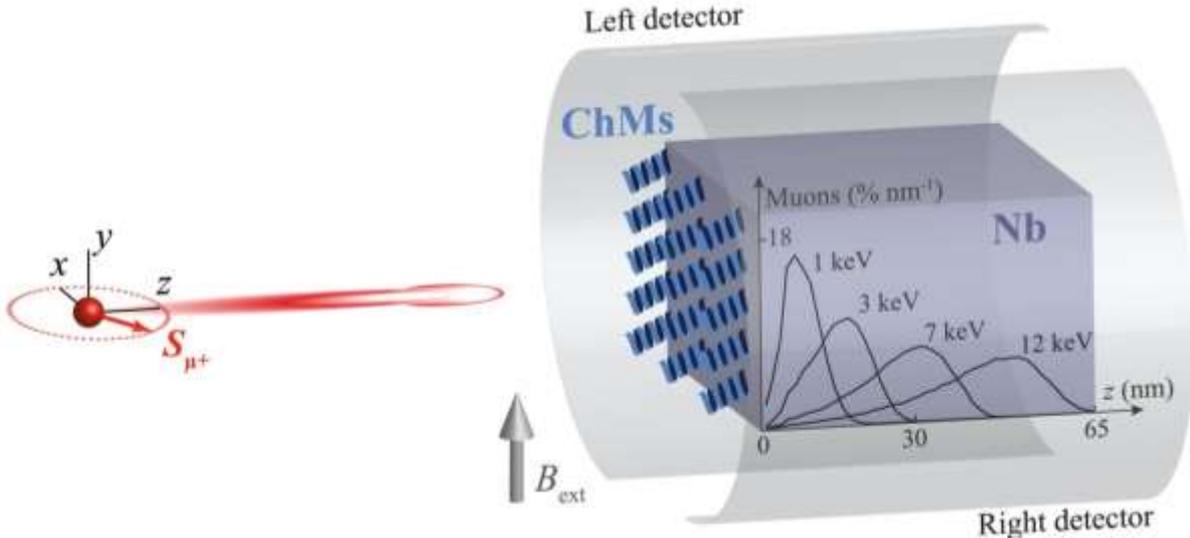

FIG. 1. Schematic of the LE-μSR measurement setup. LE-μSR in a transverse-field configuration with the external field $B_{ext}$ applied parallel to the sample surface and perpendicular to the muon's spin precession plane. Normalized muon stopping distributions simulated for the ChMs/Nb (65 nm) system are also shown for a few representative energies (black lines).

We perform the LE-μSR measurements on the ChMs/Nb (65 nm) and bare Nb (65 nm) in a transverse-field (TF) configuration, where the external field, $B_{ext}$, is applied parallel to the sample surface (*y*-axis in Fig. 1) and perpendicular to the precession plane of the muon's spin (*xz*-plane in Fig. 1). At a given implantation energy $E$, the muons' spins precess around $B_{loc}$ at an average Larmor frequency $\bar{\omega}_s(E) = \gamma_\mu \bar{B}_{loc}(E) = \int \gamma_\mu B_{loc}(z) p(z,E) dz$, where $\gamma_\mu$ = 851.616 Mrad (sT)$^{-1}$ is the muon's gyromagnetic ratio and $p(z,E)$ is the muon stopping distribution [25] simulated with the Monte Carlo algorithm TrimSP [26]. We reconstruct the depth dependence of the local field by determining $\bar{B}_{loc}(\bar{z}(E))$, meaning the average local field experienced by muons at different implantation energies, since each implantation energy $E$ corresponds to an average implantation depth $\bar{z}(E)$ – which we determine from $p(z, E)$ (black profiles in Fig. 1).

The signal that we measure, known as asymmetry $A_S(t, E)$, it is proportional to the muon spin polarization, and it is experimentally obtained from the difference in the number of muons' decaying events counted by the left and right arrays of positron detectors (Fig. 1) normalized to the total counts of the detectors. At a given $E$, $\bar{B}_{loc}(\bar{z}(E))$ is obtained from a single-energy fit of $A_S(t, E) \propto e^{-\frac{\bar{\sigma}^2 t^2}{2}} \cos[\gamma_\mu \bar{B}_{loc} t + \varphi_0(E)]$, where $\bar{\sigma}$ and $\varphi_0$ are the muons' depolarization rate and average initial precession phase, respectively, which we also extract from the fit [27].

Fig. 2(a)-(b) show $\bar{B}_{loc}(\bar{z}(E))$ measured for both Nb (65 nm) samples in the normal state, at $T$ = 10 K, and in the superconducting state at $T$ = 2.8 K. The superconducting critical temperature $T_c$ is ~ 9.15 K for the Nb (65 nm) sample, as shown in Appendix B (typically $T_c$ is reduced by less than 1% upon ChMs adsorption [28]). In the normal state, $\bar{B}_{loc}$ is independent on depth and represents an accurate *in situ* measurement of the applied field $B_{ext}$ ~ 302.6 Gauss. In the superconducting state, our measurements show that $\bar{B}_{loc}$ exhibits a significant change with respect to $B_{ext}$ for both the bare Nb and the ChMs/Nb samples, as a result of the superconducting screening currents flowing inside the Nb films. Nevertheless, we also observe a pronounced difference in the $\bar{B}_{loc}$ profiles at 2.8 K between the two samples, although the two Nb thin films are grown in the same conditions (Fig. 2(a); filled symbols).

The LE-µSR data in Fig. 2(a) show that for the bare Nb sample, $\bar{B}_{loc}$ indeed follows the profile expected for a slab of conventional S of thickness $d_s$ with a $B_{ext}$ applied parallel to its surface [29]. The ChMs/Nb sample instead shows a $\bar{B}_{loc}$ profile exhibiting an enhancement (of several Gauss) in diamagnetic screening close to the ChMs/Nb interface with a simultaneous suppression in diamagnetism (i.e., a paramagnetic shift) deeper inside Nb (at $E$ > 5 keV). The crossover between these two regions is reported in detail in Fig. 2(b). We verify the presence and absence of ChMs in the ChMs/Nb and bare Nb samples of Fig. 2, respectively, by performing X-ray photoelectron spectroscopy measurements after the completion of the LE - µSR experiment (see Appendix B). To collect the data shown in Fig. 2, before each energy scan below $T_c$, we degauss the magnet of the LE – µSR setup above $T_c$ (at $T$ ~ 200 K). The magnet has a remanent field of ~0.3 Gauss (measured using muons as accurate magnetic field sensors), which is an order of magnitude smaller than the difference between the $\bar{B}_{loc}$ profiles for ChMs/Nb and Nb in Fig. 2(a). The measurement protocol followed, and the magnitude of the remanent field therefore rule out trapped magnetic flux or pinned vortices as possible explanation for the unconventional Meissner screening observed for ChMs/Nb.

The change in the screening properties of the Nb thin film upon the adsorption of ChMs is further evidenced by the variation in the muon depolarization rate, $\bar{\sigma}$, which is related to the field distribution width and therefore to the homogeneity in the local field experienced by muons at their implantation sites. In the normal state, where the screening currents are absent and muons primarily experience the dipolar fields of the Nb nuclear moments, $\bar{\sigma}$ is the same for both ChMs/Nb and bare Nb, as expected (Fig. 2(c); hollow symbols). In the superconducting state, however, although the trends of $\bar{\sigma}$ as a function of $E$ are qualitatively the same for both the ChMs/Nb and Nb samples, we observe a clear shift in amplitude between them (Fig. 2(c); filled symbols). We also note that the shift between the two $\bar{\sigma}(E)$ profiles becomes more pronounced in the same $E$ range (2 - 8 keV; Fig. 2(d)), where the crossover between the two $\bar{B}_{loc}$ profiles (Fig. 2(b)) occurs.

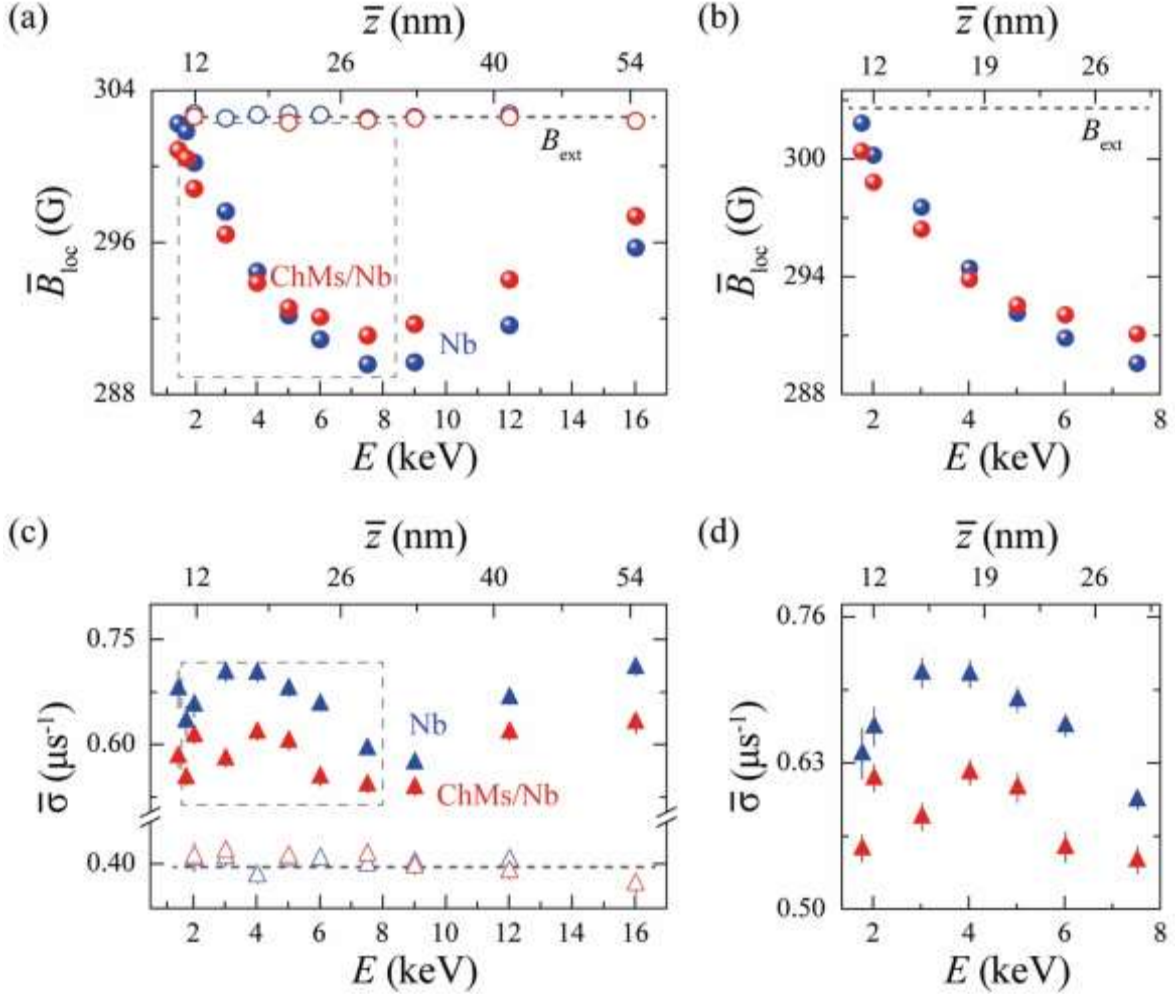

FIG. 2. Magnetic field response in ChMs/Nb and bare Nb probed by LE-μSR. Average local magnetic field $\bar{B}_{loc}$ (a)-(b) and muon spin depolarization rate $\bar{\sigma}$ (c)-(d) as a function of implantation energy $E$ (bottom axes) and average muon stopping depth $\bar{z}$ (top axes). The error bars in (a)-(b) are within the size of the symbols. The data are collected for both samples in the normal state at 10 K (hollow symbols; blue for Nb (65 nm) and red for ChMs/Nb (65 nm)) and in the superconducting state at 2.8 K (filled symbols; blue for Nb (65 nm) and red for ChMs/Nb (65 nm)). The data in (b) and (d) correspond to those in the dashed boxes in (a) and (c).

The data obtained from the single-energy asymmetry fits in Fig. 2 clearly demonstrate a modification of the screening properties of the S film upon adsorption of ChMs up to tens of nanometers away from the ChMs/Nb interface. However, the $\bar{B}_{loc}$ profiles in Fig. 2(a)-(b) include the contribution of depth averaging due to the width of the muon stopping distributions $p(z, E)$ (see Fig. 1). To obtain a more accurate local field profile inside the samples, a global fit is commonly used, where the experimental data obtained from all energies are combined together and fitted to an analytical model for $B_{loc}(z)$, as further discussed below.

## III. EVIDENCE FOR MAGNETIC SPIN ACTIVITY OF ChMs

The results presented in Fig. 2 suggest that ChMs, although these are non-magnetic in solution, act as a spin-active layer once adsorbed onto the surface of a S. This assumption, for which indirect evidence is experimentally demonstrated in ref. [12] and theoretically predicted in ref. [18], is at the heart of the theoretical model and corresponding analytical expressions used for the global fit presented below. To directly validate the spin activity of the SAM of ChMs, we perform two different experiments based on LE-µSR.

In the first experiment, we check the effect of reversing the direction of the in-plane applied external magnetic field $B_{ext}$ on an additional ChMs/Nb (65 nm) sample, different from that reported in Fig. 2. Fig. 3 shows the normalized shift in $\bar{B}_{loc}$ measured for this ChMs/Nb sample for two opposite in-plane $B_{ext}$ values. The normalized shift is calculated as $(\bar{B}_{loc, 2.8\,K} - \bar{B}_{loc, 10\,K})/\bar{B}_{loc, 10\,K}$, where $\bar{B}_{loc, 2.8\,K}$ is the local field measured below $T_c$ at $T = 2.8$ K, whereas $\bar{B}_{loc, 10\,K}$ is the local field measured above $T_c$ at $T = 10$ K. The data in Fig. 3 demonstrate that, whilst approaching the ChMs/Nb interface (i.e., for $E < 4$ keV), a gap in the $\bar{B}_{loc}$ shifts progressively between the two field orientations. In particular, $\bar{B}_{loc}$ measured in $B_{ext} = + 300$ Gauss exhibits an enhancement in the diamagnetic shift from the normal-state $\bar{B}_{loc}$ at the ChMs/Nb interface (consistently with data for the other ChMs/Nb sample in Fig. 2), whilst $\bar{B}_{loc}$ measured in $B_{ext} = - 300$ Gauss almost recovers the normal-state $\bar{B}_{loc}$ value at $E = 1.6$ keV.

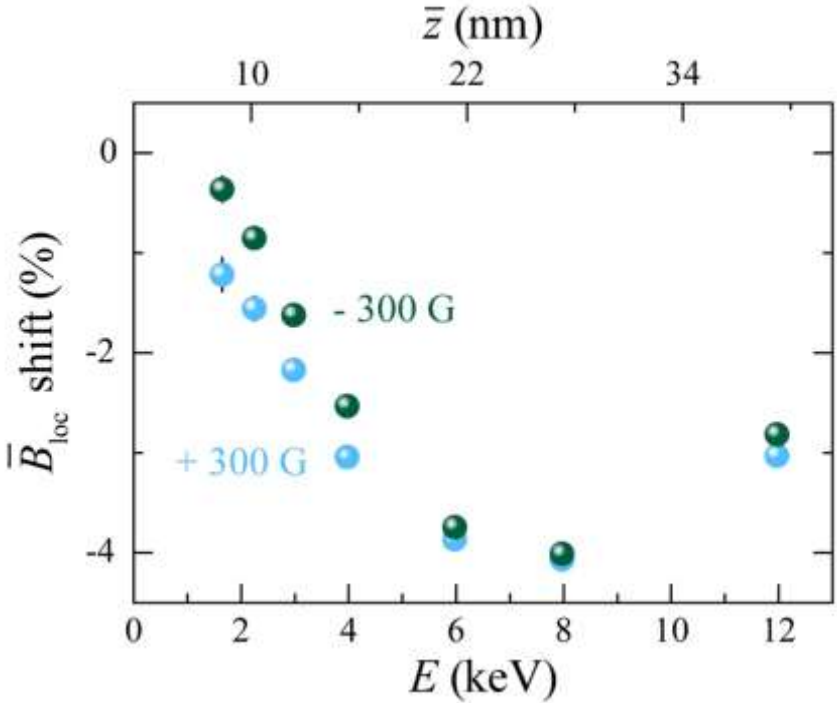

FIG. 3. Magnetic dependence of the unconventional Meissner effect in ChMs/Nb. Shift in the average local magnetic field $\bar{B}_{loc}$ between the superconducting and normal state in a ChMs/Nb (65 nm) sample as a function of muon implantation energy $E$ (bottom axis) and average muon stopping depth $\bar{z}$ (top axis) for $B_{ext} \sim 300$ Gauss (light blue symbols) and $B_{ext} \sim -300$ Gauss (green symbols). The shift in $\bar{B}_{loc}$ is determined as the difference between $\bar{B}_{loc}$ at $T = 2.8$ K and $\bar{B}_{loc}$ at $T = 10$ K normalized to $\bar{B}_{loc}$ at $T = 10$ K.

At higher implantation energies, meaning moving away from the ChMs/Nb interface, the difference between the $\bar{B}_{loc}$ shifts in Fig. 3 nearly vanishes, which is consistent with the fact that the unconventional Meissner screening measured in ChMs/Nb originates from a superconducting proximity effect at the ChMs/Nb interface.

The different dependence of $\bar{B}_{loc}$ in the superconducting state for the two field directions shown in Fig. 3 also confirms that the SAM of ChMs is spin-polarized, with a net spin component that can be aligned either along, or opposite to, the applied field, and which in turn results in a suppression or enhancement of the local magnetic field near the ChMs/Nb interface. This *collective* behavior of ChMs, which breaks in-plane rotational symmetry, has been experimentally observed upon adsorption ChMs to a F layer [30] and has been theoretically predicted to arise as result of the magnetic exchange interaction between ChMs [31]. The dependence of the unconventional Meissner screening effect on the relative orientation of the net polarization of the ChMs about $B_{ext}$ also supports our theoretical model, which assumes a spin-activity of the ChMs layer, as discussed below. Importantly, the data sets for the two field directions in Fig. 3 are acquired on the same ChMs/Nb sample, after the same cool down. This result rules out any sample-to-sample variations in the superconducting or physical properties of the Nb thin films as an explanation for the difference in the local field profiles reported in Fig. 2. Also, we note the data in Fig. 2 are collected after zero-field-cooling the ChMs/Nb sample, whereas the data reported in Fig. 3 for the same field $B_{ext}$ = +300 Gauss are collected after field-cooling the ChMs/Nb sample, which shows that the enhancement in Meissner screening at the ChMs/Nb interface is consistently observed in $B_{ext}$ = +300 Gauss independently on the field-cooling history.

To further corroborate the spin activity of the ChMs, we perform a second experiment where we carry out LE-μSR measurements in zero field (ZF) on an additional ChMs/Nb sample with Nb thickness of ~ 55 nm and $T_c$ ~ 8.7 nm, and also on a bare Nb (55 nm) film prepared in the same deposition run under identical conditions (Fig. 4). The ZF measurements are carried out for both samples at $E$ = 8 keV (corresponding to $\bar{z}$ ~ 30 nm, i.e., approximately to the middle of the Nb films) and $E$ = 3 keV (closer to the surface) in the temperature range from 2.8 K to 10.5 K. We fit the $A_S(t,E)$ signal measured in ZF to the Kubo-Toyabe function [33] (see Appendix C), $P_z(t) = e^{-\bar{v}t}\left(\frac{1}{3} + \frac{2}{3}(1-\bar{\sigma}^2 t^2)\exp\left[-\frac{\bar{\sigma}^2 t^2}{2}\right]\right)$, in which $\bar{\sigma}$ is associated with the depolarization of muons due to nuclear moments and other static dipolar moments in the sample, while $\bar{v}$ accounts for contributions to the muons' depolarization arising from the appearance of additional small magnetic field fluctuations in the sample.

Figure 4 shows the temperature dependence of $\bar{\sigma}$ in both the ChMs/Nb (55 nm) and Nb (55 nm) samples measured at $E$ = 3 keV. We observe that $\bar{\sigma}$ for the ChMs/Nb sample is larger than that for the Nb sample at all $T$s, but the difference between the $\bar{\sigma}$ values of the two samples increases systematically only below $T_c$ (see inset in Fig. 4). The overall increase in $\bar{\sigma}$ with decreasing $T$ for each sample can be related to temperature-dependent diffusion of muons in the presence of impurities in Nb thin films, even though in the presence of impurities it has been observed that $\bar{\sigma}$ should flatten or decrease for $T$ below $T_c$ [33-34]. Although $\bar{\sigma}$ indeed flattens for the bare Nb sample below $T \sim 6.5$ K consistently with previous studies [33-34], for the ChMs/Nb samples we observe that, in the same $T$ range, $\bar{\sigma}$ increases with reducing $T$. Considering that both Nb samples were grown in the same deposition and are thus expected to have similar impurity and defect densities, we infer that the relative change in $\bar{\sigma}$ between ChMs/Nb and Nb and its occurrence right below $T_c$ can only be explained as due to the spin activity of the SAM of ChMs.

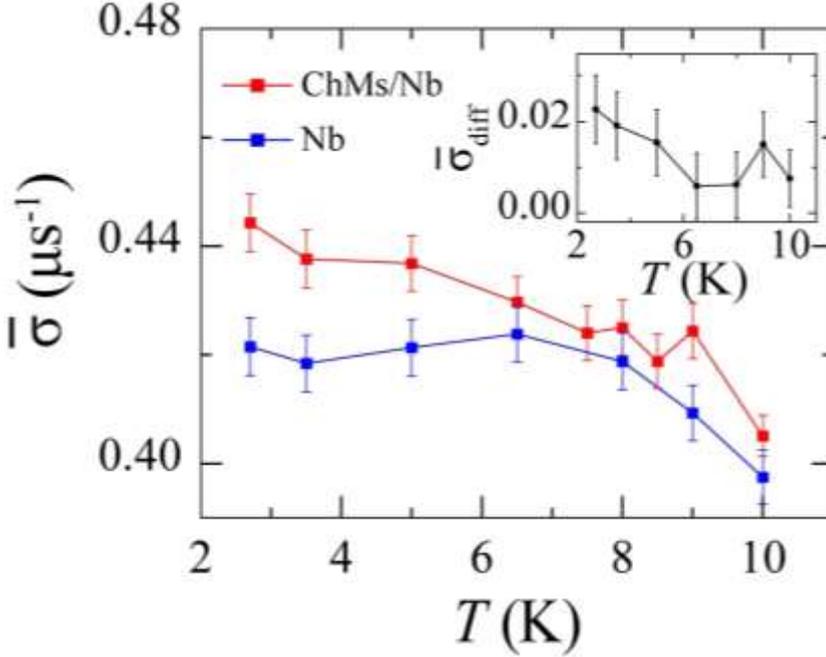

FIG. 4. Zero-field LE-μSR in ChMs/Nb and bare Nb. Average muon depolarization rate $\bar{\sigma}$ due to measured in ZF at $E = 3$ keV across the superconducting transition for a ChMs/Nb (55 nm) sample (red curve) and a bare Nb (55 nm) sample (blue curve). The inset shows the difference in $\bar{\sigma}$ ($\bar{\sigma}_{\text{diff}}$) measured between the two samples as a function of temperature.

The same behavior is observed also deeper inside the sample, for implantation energy of at $E = 8$ keV (see Fig. 10 (a)), although the difference in $\bar{\sigma}$ between the two samples at $E = 8$ keV starts increasing at somewhat lower $T$s than those reported in Fig. 4 for $E = 3$ keV, which also confirms that the origin of the effect must be related to dipolar fields originated by the adsorbed ChMs at the ChMs/Nb interface. The ZF LE-μSR data therefore support the results on the asymmetry of the unconventional Meissner screening with respect to the $B_{\text{ext}}$ direction reported in Fig. 3 and show that a SAM of ChMs acts as a magnetically active layer – which the superconducting condensate "senses" below $T_c$, inducing local screening and therefore an increase in $\bar{\sigma}$.

We also perform transverse-field measurements on the same ChMs/Nb (55 nm) sample to which the zero-field data in Fig. 4 refer. The results of these transverse-field measurements are reported in Appendix E and show a local field profile consistent with that measured for the two other ChMs/Nb samples with 65-nm-thick Nb reported in Fig. 2 and Fig. 3, which demonstrates the reproducibility of the unconventional Meissner screening in Nb thin films upon ChMs adsorption.

## IV. ANALYSIS AND DISCUSSION

To determine a suitable analytical expression for the global fit of the LE-μSR data in Fig. 2, we first numerically calculate the field profile expected theoretically in ChMs/Nb. Based on the ZF results in Fig. 4, we model the ChMs/Nb system as an insulating spin-active layer (the ChMs) coupled to a S (Nb). Using the appropriate boundary conditions for a spin-active layer, having a net magnetization component opposite to the applied field direction, we apply the Green's function formalism and solve the quasi-classical Usadel equation simultaneously with the Maxwell equations, through nested self-consistency iterations, in order to determine the spatial variations of the superconducting OP, denoted as Δ, and the magnetic field profile inside ChMs/Nb (see Appendix D). The insulating nature of ChMs, which we assume for the numerical model, is confirmed by previous transport measurements on devices where ChMs are used as an interlayer between two metallic electrodes [12].

The results of our numerical calculation of the field profile are shown in Fig. 5(b). We find that the SAM of ChMs has a two-fold effect on varying the superconducting screening inside Nb, which results in a field profile (yellow curve in Fig. 5(b)) that differs from that expected for conventional Meissner screening (dashed black curve in Fig. 5(b)). First, our theoretical model suggests that the superconducting proximity effect between the spin-active layer of ChMs and Nb leads to the generation of odd-frequency spin-triplet pairs, which in turn induce a net magnetization $M$ inside Nb (red area in Fig. 5(b)). This contribution appears in addition to the diamagnetic screening that is already present in Nb due to conventional spin-singlet pairs. In our model, we consider only the generation of spin-triplet pairs with spin-projection $S = 0$ along the magnetization direction of the spin-active interface and do not include fully polarized spin-triplet pairs with $S = \pm 1$, since $S = 0$ pairs are sufficient to generate an unconventional Meissner response due to their odd-frequency nature [8,35,36]. Second, our theoretical analysis suggests the magnetization stemming from the ChMs leads to a suppression of Δ at the ChMs/Nb interface compared to its value $\Delta_0$ for the bare Nb film (blue area in Fig. 5(b)), in analogy with what is expected for the interface between a S and an insulating F. According to our simulations, both Δ and $M$ decay on a length scale of the order of the Nb superconducting coherence length $\xi_s$ which is of ~ 13-16 nm for Nb thin films with similar properties [37]. The increase in diamagnetic screening at the Nb surface for the ChMs/Nb sample with respect to the bare Nb sample which we find experimentally results within our model from the aforementioned boundary condition, meaning that the molecular layer obtains a net magnetization component antiparallel to the applied field direction. The results of reversing the field direction in Fig. 3 are also consistent with this model's assumption because, upon reversing $B_{\text{ext}}$, the net magnetization generated by the ChMs layer becomes aligned parallel (rather than antiparallel) to $B_{ext}$, which leads to a positive shift in $B_{\text{loc}}$ at the surface.

To perform the global fit mentioned above, we need to derive an analytical expression that can properly represent the theoretical field profile calculated numerically. For the bare Nb sample we use the analytical expression for conventional Meissner screening, here denoted by $B_{\text{loc, conv.}}(z)$, given by the Ginzburg-Landau theory for a superconducting slab with thickness $d_s$ in the presence of an external field $B_{\text{ext}}$ applied parallel to its surface [29],

$$B_{\text{loc, conv.}}(z) = B_{\text{ext}} \cosh\left(\frac{z - \frac{d_s}{2}}{\lambda_{\text{GL}}}\right) / \cosh\left(\frac{d_s}{2\lambda_{\text{GL}}}\right), \quad (1)$$

where $\lambda_{GL}$ is the Ginzburg-Landau penetration depth that we use as a fitting parameter. The blue curve in Fig. 5(a) is derived by convoluting the $B_{\text{loc, conv.}}(z)$ obtained from the global fit (yielding $\lambda_{GL} = 66.4$ ($\pm$ 1.2) nm) with the $p(z, E)$ distributions (see Appendix E). This curve reproduces the LE-µSR data from single-energy fits (Fig. 5(a); blue symbols), thus confirming that the bare Nb sample indeed exhibits conventional Meissner screening.

For the ChMs/Nb sample, we adopt an analytical expression for the local field, here named $B_{\text{loc, unc.}}(z)$, which captures the two physical effects described by our theoretical field profile and underlying the unconventional screening in ChMs/Nb. We set

$$B_{\text{loc, unc.}}(z) = B_{\text{Meiss, unc.}}(z) + B_{\text{spin}}(z) =$$
$$= B_{\text{ext}} \cosh\left(\frac{z - \frac{d_s}{2}}{\lambda(z)}\right) / \cosh\left(\frac{d_s}{2\lambda(z)}\right) + B_{\text{spin}} \exp\left(-\frac{z}{\xi_{\text{spin}}}\right), \quad (2)$$

where the term $B_{\text{Meiss, unc.}}(z)$ is a modification of $B_{\text{loc, conv.}}(z)$ obtained by introducing a depth dependence for $\lambda(z)$ to mimic the suppression of $\Delta$ from the ChMs/Nb interface. The additional term $B_{\text{spin}}(z) = B_{\text{spin}} \exp\left(-\frac{z}{\xi_{\text{spin}}}\right)$ is introduced in $B_{\text{loc, unc.}}(z)$ to model the *M* contribution inside Nb due to the generation of spin triplets. $B_{\text{spin}}(z)$ mimics the effect of *M*, and indeed our theoretical simulations suggest that its decay length $\xi_{\text{spin}}$ is $\sim \xi_s$.

Following reports on other LE-µSR studies [20,42], in the global fit we also assume the existence of a "dead layer" of thickness $z^* \sim 10$ nm from the Nb surface (i.e., $z = 0$; Fig. 1), where an inhomogeneous distribution of screening currents results in a reduced deviation of $B_{\text{loc}}$ from $B_{\text{ext}}$ (see Appendix E). We assume the presence of a dead layer also at the Nb/SiO$_2$ substrate interface. Muons get implanted into the SiO$_2$ substrate at $E \geq 16$ keV as suggested by the $p(z, E)$ in Appendix E and confirmed by the deviation at $E = 16$ keV of the measured field in Fig. 5(a) from the theoretical profiles for both $B_{\text{loc, conv.}}(z)$ and $B_{\text{loc, unc.}}(z)$ shown in Fig. 5(b).

Our simulations suggest that the spatial variation of $\lambda(z)$ in $B_{\text{Meiss, unc.}}(z)$ is negligible for $z > z^*$, and therefore we use a constant $\lambda_{GL}$, in addition to $B_{\text{spin}}$ and $\xi_{\text{spin}}$ as fitting parameters for $B_{loc, unc.}(z)$ (for details about the fitting procedure, see Appendix E). The global fit yields $B_{\text{spin}} = -6.6$ ($\pm$ 1.1) Gauss, and $\lambda_{GL} = 73.9$ ($\pm$ 1.0) nm which is close to $\lambda_{GL} = 66.4$ ($\pm$ 1.2) nm obtained for $B_{\text{loc, conv.}}(z)$ in bare Nb. The difference between $\lambda_{GL}$ for Nb and ChMs/Nb may be a combination of a small sample-to-sample deviation as well as of the fact that a larger $\lambda_{GL}$ mimics the suppression in $\Delta$ at the ChMs/Nb interface. The fit also returns a length scale for the exponential decay of $B_{\text{spin}}$ equal to $\xi_{\text{spin}} = 6.0$ ($\pm$ 2.0) nm. The effect of $B_{\text{spin}}$ should also extend to part of the dead layer since in reality spin triplets must be present herein, but this effect cannot be resolved by muons due to the inhomogeneous current distributions which they experience inside the dead layer. Given the value obtained for $\xi_{\text{spin}}$ and $z^*$ we conclude that the effect of $B_{\text{spin}}$ extends over a length scale comparable to $\xi_s$, in agreement with our theoretical model.

The field profile, which is obtained by convoluting $B_{\text{loc, unc.}}(z)$ generated by the global fit with the muon stopping distributions $p(z, E)$, it is represented by the red curve in Fig. 5(a) and it properly reproduces the experimental data from the single-energy asymmetry fits (Fig. 5(a), red symbols). This result confirms that our theoretical model and its analytical description can indeed explain and account for the physical effects underlying the unconventional Meissner screening inside Nb induced by ChMs through the generation of odd-frequency spin-triplet states. We also note that the odd-frequency nature of the spin-triplet states generated at the ChMs/Nb interface is an essential condition for the observation of an unconventional Meissner response, as already demonstrated by previous theoretical [35] and experimental [8, 24] studies. The combination of inversion symmetry breaking at the ChMs/Nb interface with the high spin-orbit coupling in Nb can lift Kramers degeneracy and induce a mixed superconducting state with spin-singlet and spin-triplet pairs that are even in frequency, but such even-frequency pairs are not sufficient to generate an unconventional Meissner response, as also evidenced by beta-nuclear magnetic resonance measurements (β-NMR) in $NbSe_2$ [43], a S material that should host such a mixed even-frequency spin-singlet/spin-triplet state.

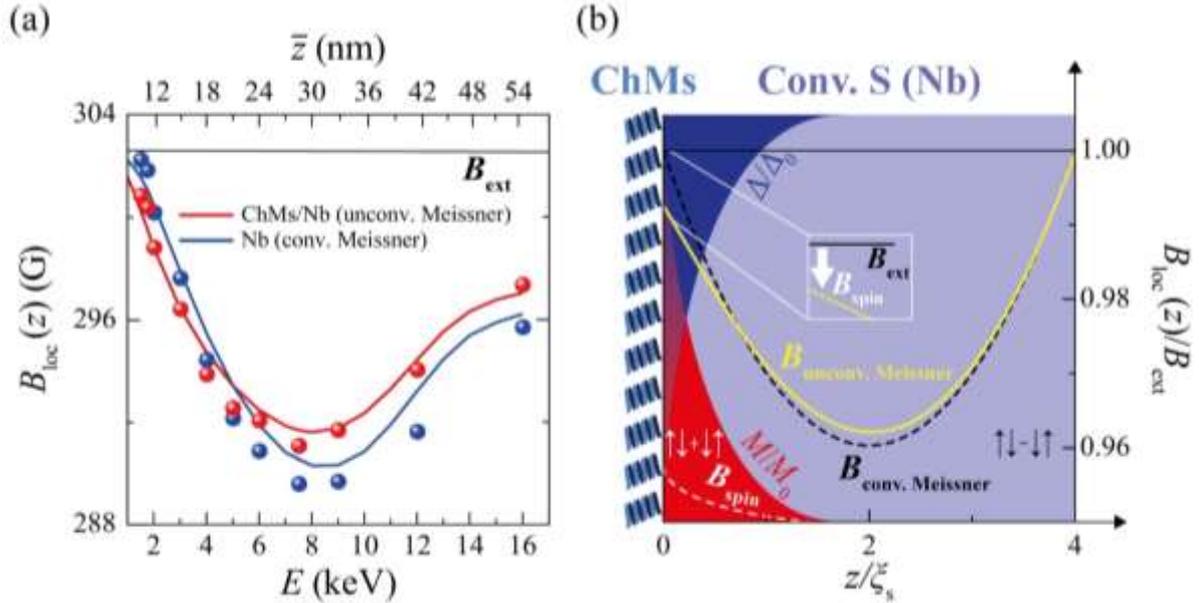

FIG. 5. Global fit and theoretical model of the local field profile in ChMs/Nb. (a) $B_{\text{loc}}(z)$ profile from the global fit convoluted by the muon stopping distributions (solid lines) and comparison to single-energy asymmetry data (filled symbols) for bare Nb based on the Ginzburg-Landau model (blue solid curve), and for the ChMs/Nb sample based on the theoretical field profile for unconventional Meissner screening due to spin triplets (red solid curve). (b) Theoretical field profile for unconventional Meissner screening in ChMs/Nb (yellow curve) calculated by solving simultaneously the Usadel and Maxwell equations. This field profile differs from that expected for conventional Meissner screening based on Ginzburg-Landau model (black dashed curve) because of two additional terms, both decaying on a length scale comparable to the Nb coherence length $\xi_S$: one term ($B_{\text{spin}}$) related to the generation of spin triplets which leads to a magnetization $M$ component adding to the magnetization $M_0$ inside Nb (red area) and a second term related to the suppression of $\Delta$ at the ChMs/Nb interface (blue area).

In conclusion, the magnetic activity of ChMs is responsible for the generation of odd-frequency spin-triplet correlations, for a suppression of the superconducting gap, and for inducing a spin-magnetization in Nb (Fig. 5(b)), which in turn modify the screening current distribution over a distance of the order of $\xi_s$ inside Nb. Our results demonstrate that a single monolayer of ChMs constitutes a spin-active layer that can radically modify the screening properties of a conventional thin film S not only on a local scale limited to the ChMs/S interface, but much deeper inside the S by adding an odd-frequency spin-triplet component to the superconducting state. In addition, by varying the direction of the applied external field with respect to the spin-polarization induced by the ChMs, we can tune the superconducting screening properties of the ChMs/Nb system. Our findings therefore pave the way for the fabrication of hybrid molecular-superconducting devices for superconducting logic and memory operations, where the magnetic flux coupled through a S can be not only varied in a local controlled way through the adsorption of ChMs onto selected areas of the superconductor surface, but also modulated by switching the orientation of an applied external magnetic field.

## ACKNOWLEDGMENTS

A.D.B. and R.H. acknowledge funding from the Humboldt Foundation in the framework of a Sofja Kovalevskaja grant endowed by the Alexander von Humboldt foundation and, along with E.S., funding from the Deutsche Forschungsgemeinschaft through the SPP priority program 2244. O. M. and Y. P. acknowledge funding from the Niedersachsen Ministry of Science and Culture and from the Academia Sinica, Hebrew University Research Program. O.M. also acknowledges support through the Harry de Jur Chair in Applied Science. J. L. and M.A. acknowledge funding from the Research Council of Norway through its Centres of Excellence scheme (project number 262633, QuSpin). The muSR measurements were performed at the Swiss Muon Source (SμS), at the Paul Scherrer Institute in Villigen, Switzerland.

# APPENDIX A: MATERIALS AND METHODS

## 1. Sample preparation

The Nb thin films are grown onto $SiO_2$ substrates by direct current (dc) magnetron sputtering in an ultra-high vacuum deposition chamber with a base pressure below $10^{-8}$ torr. The pairs of Nb samples with a specific thickness (i.e., 65 nm or 55 nm), of which one is used for the adsorption of ChMs and the other as control sample (i.e, without ChMs) for the LE-μSR measurements, are grown in the same deposition run. The Nb samples with different thicknesses instead are grown using the same growth parameters and conditions but in different deposition runs. The electrical resistance of the thin films is measured in a four-probe configuration inside a cryogen-free system (Cryogenic Ltd.) with base temperature of ~ 1.5 K using a current-bias setup with current equal or less than 0.1 mA. The low-angle X-ray reflectometry measurements on the Nb thin films are performed using a Rigaku Smartlab diffractometer with a double-bounce channel cut Ge (220) monochromator and an incident slit of 0.05 mm.

The chiral molecules used are alpha helix poly alanine molecules, AHPA, with 36 amino acids CAAAAKAAAAKAAAAKAAAAKAAAAKAAAAKAAAAK (C stands for cysteine, A for alanine and K for lysine) and with a calculated length of 5.4 nm (molecules are produced by Sigma Aldrich). A self-assembled monolayer (SAM) of such molecules is adsorbed onto Nb after dipping the Nb samples for 12 hours into a 1 mM solution of chiral molecules in ethanol, followed by rinsing in ethanol and drying under $N_2$ flow inside a glove box with $N_2$ atmosphere. To ensure reproducibility between samples, we also follow the same preparation steps for the bare Nb thin films used for the LE-μSR measurements, meaning that we also dip them in an ethanol bath for 12 hours inside a $N_2$ atmosphere but without ChMs.

## 2. Surface topography measurements

The KPFM measurements are conducted on an Ntegra modular apparatus (NT-MDT) embedded with a scanning probe microscopy option and using a conductive tip. The KPFM measurements are performed using tip bias and a grounded sample. A double-path measurement technique is adopted, where in the second path (i.e., the KPFM data collection) the tip is kept at a fixed distance of 10 nm from the sample surface.

## 3. X-ray photoelectron spectroscopy (XPS)

XPS measurements are done on a Kratos Axis Supra X-ray photoelectron spectrometer (Karatos Analytical Ltd., Manchester, U.K. installed in 2019) with an Al Kα monochromatic radiation source (1486.7 eV) using a 90° takeoff angle (normal to analyser). The vacuum pressure in the analysing chamber is maintained at $2\times10^{-9}$ Torr. The XPS high-resolution spectra for C 1s, O 1s, N 1s, S 2p and Nb 3d levels are obtained with pass energy 20 eV and step 0.1 eV. The binding energies are calibrated relative to the C 1s peak energy position at 285.0 eV. Data analysis of the XPS data is carried out using the Casa XPS (Casa Software Ltd.) and the ESCApe data processing software programmes (Kratos Analytical Ltd.).

## APPENDIX B: SAMPLES CHARACTERISATION

To verify adsorption of ChMs onto Nb, we define areas for selective adsorption of ChMs on our Nb thin films using electron beam lithography (EBL). Squared areas (10 μm x 10 μm in size) are patterned by EBL into a poly(methyl methacrylate), PMMA, resist layer, which is then developed leaving the Nb surface available for molecular adsorption. We adsorb ChMs by dipping the sample in solution according to the procedure described in Appendix A, and we then remove the PMMA resist by dipping the samples in acetone.

To verify the presence of the molecules in the areas defined by EBL and their absence in the PMMA areas unexposed to the electron beam, we use Kelvin probe force microscopy (KPFM). The KPFM data in Fig. 6 show that the surface potential pattern perfectly matches the lithographic pattern defined in the PMMA layer, which demonstrates the successful adsorption of ChMs onto the Nb areas defined by EBL.

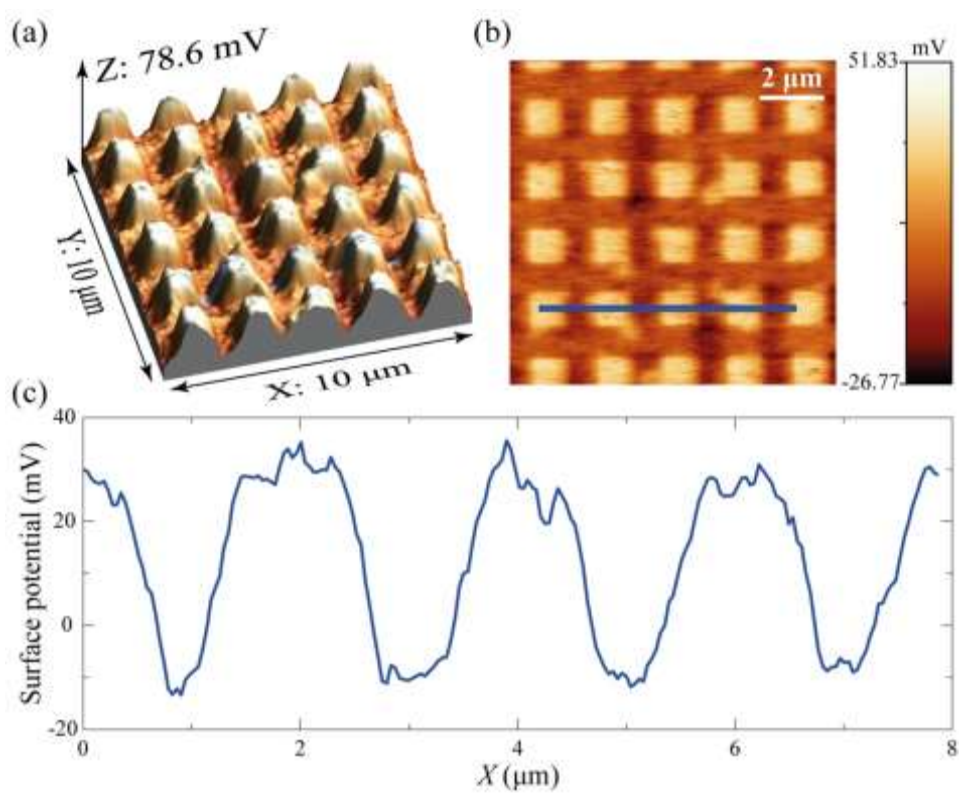

FIG. 6. Surface potential of a Nb thin film with selective adsorption of ChMs. Surface potential image in 3D (a) and 2D (b) measured by Kelvin probe force microscopy on a 10 μm x 10 μm area of a Nb (65 nm) thin film with ChMs selectively adsorbed onto its surface. The surface potential in (c) is measured along the blue line in (b). Selective adsorption of ChMs is obtained by defining squares by e-beam lithography into a PMMA resist layer and then adsorbing the ChMs onto the patterned area. The PMMA is removed after the adsorption. The presence of the ChMs in the adsorbed area is evidenced by a variation in the surface potential of ~35mV.

We also carry out XPS measurements on the same ChMs/Nb (65 nm) samples investigated by LE-μSR to confirm the presence of the ChMs on Nb during the LE-μSR experiment. We consider contributions to the XPS spectra coming from N 1s as evidence for the presence of ChMs, since N is only present in the chemical structure of the AHPA ChMs which we use in our study, but it is absent in Nb. Fig. 7 indeed shows that spectral contributions from N 1s are only detected in the ChMs/Nb (65 nm) sample investigated by LE-μSR, but these spectral contributions are absent in the bare Nb (65 nm) thin films used in the same LE-μSR experiment.

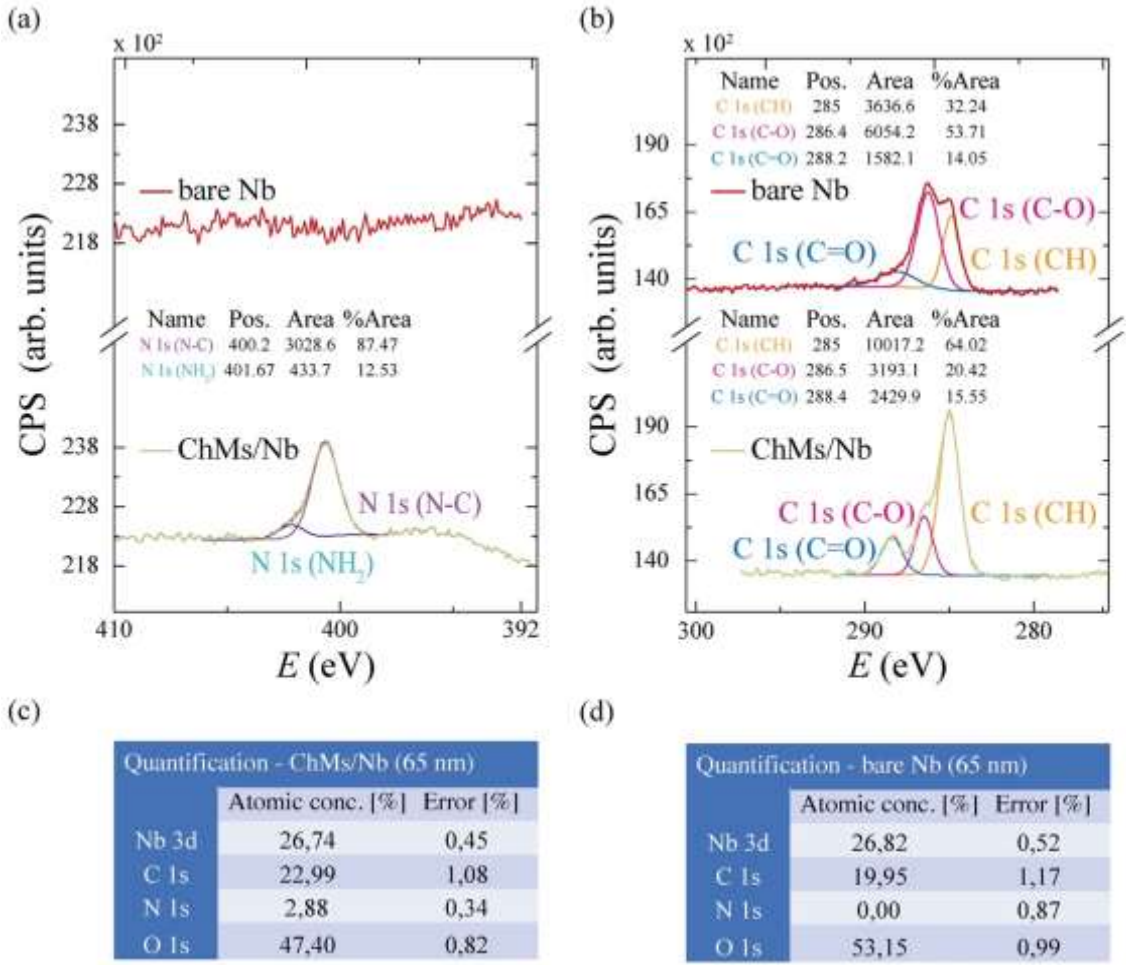

FIG. 7. X-ray photoelectron spectroscopy measurements. (a)-(b) XPS spectra for N 1s (a) and C 1s (b) for ChMs/Nb (65 nm) (green curves) and Nb (65 nm) (red curves) with corresponding analysis of the chemical bonds contributing to the spectra. The XPS spectra are collected on the same ChMs/Nb (65 nm) and Nb (65 nm) samples used for the LE-μSR measurements discussed in the main paper, after completion of the LE-μSR study. (c)-(d) Quantification of the atomic concentration for Nb, C, N, O on the surface of the same bare Nb (65 nm) (c) and ChMs/Nb (65 nm) (d) sample determined from analysis of XPS spectra acquired on the samples. The existence of the molecules on the surface of the ChMs/Nb sample is verified through the signal contribution from N1s orbitals, which are present in the AHPA molecules used in this work.

In addition to verifying the ChMs adsorption onto Nb and the presence of the ChMs in the samples investigated by LE-μSR, we also perform a characterization of the electronic transport and structural properties of the same Nb thin films used for the LE-μSR experiment.

The resistance versus temperature ($R(T)$) curve in Fig. 8 is measured in a current-biased setup with a four-point measurement technique, and it shows that the Nb films with thickness of 65 nm have a superconducting critical temperature ($T_c$) of ~ 9.15 K, and a residual resistivity ratio (RRR) of ~ 4. We also verify that the $T_c$ of the same film does not vary significantly after adsorption of the ChMs.

Low-angle X-ray reflectometry (XRR) measurements are also performed on the same Nb thin films used for the LE-µSR experiment (Fig. 9). The analysis of the XRR spectra allows to quantify structural parameters of the samples like the thickness of the $Nb_2O_5$ oxide passivation layer, the Nb thickness, and the density of both layers ($Nb_2O_5$ and Nb). A precise determination of these parameters is important for the global fit described in Appendix E, since these parameters are fed as input to the software TrimSP to calculate the muon stopping distributions, which are then used by the global fit algorithm.

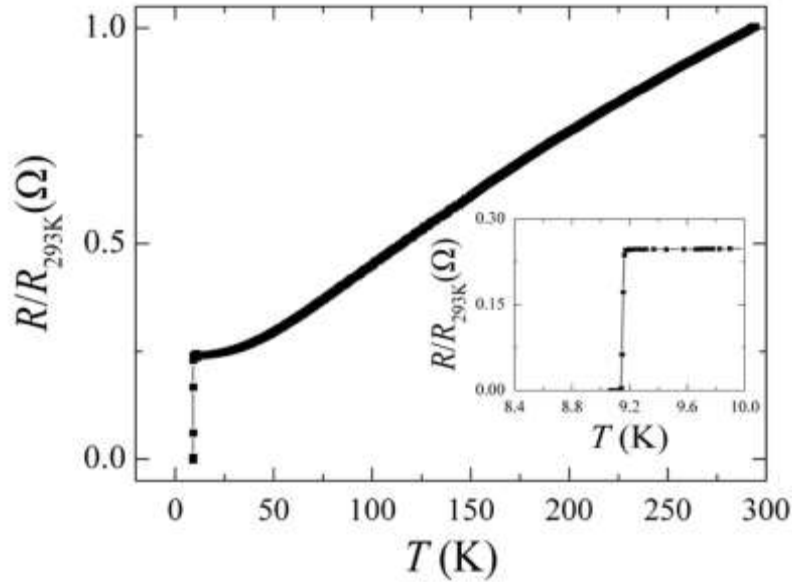

FIG. 8. Electronic transport properties of a Nb thin film used for LE-µSR measurements. $R(T)$ normalized to the resistance at 293 K ($R_{293K}$) and measured in a four-probe configuration on a Nb(65 nm) thin film, with the inset showing a zoom on the same $R(T)/R_{293K}$ curve across the superconducting transition. The data show that the thin film has $T_c \sim 9.15$ K and a RRR of ~ 4.

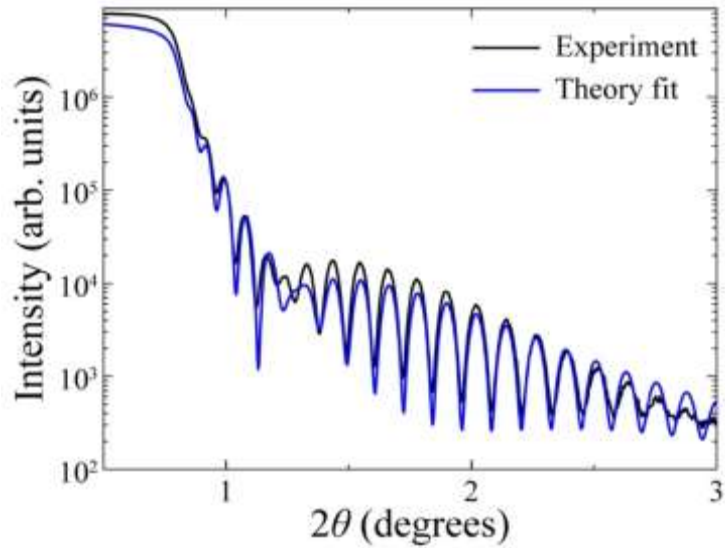

FIG. 9. X-ray reflectometry for a Nb (65 nm) thin film sample. X-ray reflectometry data (blue curve) measured on a Nb (65 nm) thin film deposited on a $SiO_2$ substrate and used for the LE-µSR measurements done to acquire the results shown in Fig. 2. The fit to the experimental data indicates the presence of a surface native oxide $Nb_2O_5$ layer with thickness of 3.847 nm $\pm$ 0.04 nm, density = 3.482 $\pm$ 0.04 g/cm$^3$ and roughness 0.796 nm $\pm$ 0.1 nm on top of a Nb thin film with thickness 64.678 nm $\pm$ 0.05 nm, density = 8.5 $\pm$ 0.04 g/cm$^3$ and roughness of 1.56 $\pm$ 0.1 nm. The density and the thickness values obtained from this fit for the $Nb_2O_5$ and Nb layers are given as input to the program TrimSP to simulate the proper muon implantation distributions for the sample.

# APPENDIX C: ZERO-FIELD LE-µSR DATA

We explain our experimental LE-µSR results based on a theoretical model that assumes that the layer of ChMs act as a spin active interface that induces the generation of odd-frequency spin-triplet pairs inside Nb. Given the lack of any direct previous experimental evidence for the magnetic spin activity of ChMs upon adsorption on Nb, we first examine the effect of reversing the applied field direction $B_{ext}$ in the transverse field (TF) setup. As shown in Fig. 3, reversing $B_{ext}$ leads to a variation of the local field profile depending on the relative alignment of the spin polarization induced in the superconducting state of Nb by the ChMs with $B_{ext}$. This result suggests a spin activity associated with the monolayer of ChMs upon adsorption onto a S layer.

To further prove the validity of our assumption on the spin activity of ChMs, we perform LE-µSR in zero field (ZF). The asymmetry function $A_s(t)$ in ZF is fitted using the Kubo-Toyabe function, which accounts for the muon depolarization induced by the Nb nuclei moments and other static moments inside the sample, times an exponentially decaying term that depends on fluctuations inside the sample. These two contributions are related to the parameters $\bar{\sigma}$ and $\bar{\nu}$, respectively, which are extracted from the fits of $A_s(t)$ based on the Kubo-Toyabe model (see Appendix E).

Fig. 10(a) shows $\bar{\sigma}$ as a function of temperature ($T$) across $T_c$ for both ChMs/Nb (55 nm) and the bare Nb (55 nm), measured at an additional energy ($E$ = 8 keV) compared to that reported in Fig. 4 of the main text. As explained in detail in Appendix E, we assume that $\bar{\nu}$ is a $T$-independent fitting parameter, meaning a parameter common to all the measurements performed in a given sample as a function of $T$. Consistently with the data in Fig. 4 of the main text measured at $E$ = 3 keV, Fig. 10(a) also shows that $\bar{\sigma}(T)$ increases below $T_c$ faster in the ChMs/Nb (55 nm) sample compared to the bare Nb (55 nm) sample.

Fig. 10(b) shows representative profiles of $A_s(t)$ corresponding to the two data points in Fig. 10(a) obtained for ChMs/Nb (55 nm) at $T$ = 2.8 K and $T$ = 10 K (i.e., measured above and below $T_c$), and their corresponding fits based on the Kubo-Toyabe model (solid lines in Fig. 10 (b)). The difference between the fitting curves at low relaxation times between 0.1 and 4 µs in Fig. 10 (b) is properly captured by the different $\bar{\sigma}$ values obtained from the fit of $A_s(t)$ based on the Kubo-Toyabe function. The variation in $\bar{\sigma}$ across $T_c$ represents strong evidence for a spin activity of the ChMs layer as explained in the main text.

We also note that the raw data for $A_s(t)$ in Fig. 10(b) deviate at high relaxation times, which is something, however, that cannot be picked up by our numerical fits. This limitation derives from our fitting procedure that takes $\bar{\nu}$ as $T$-independent, in order to avoid the large cross-correlations which otherwise we would obtain between $\bar{\sigma}$ and $\bar{\nu}$ if they were both used as fitting parameters. Therefore, although we keep $\bar{\nu}$ as $T$-independent parameter, the raw data also suggest some small variations in $\bar{\nu}$ across the superconducting transition of the ChMs/Nb (55 nm) sample, possibly related to *spontaneous spin fluctuations* associated with the ChMs layer.

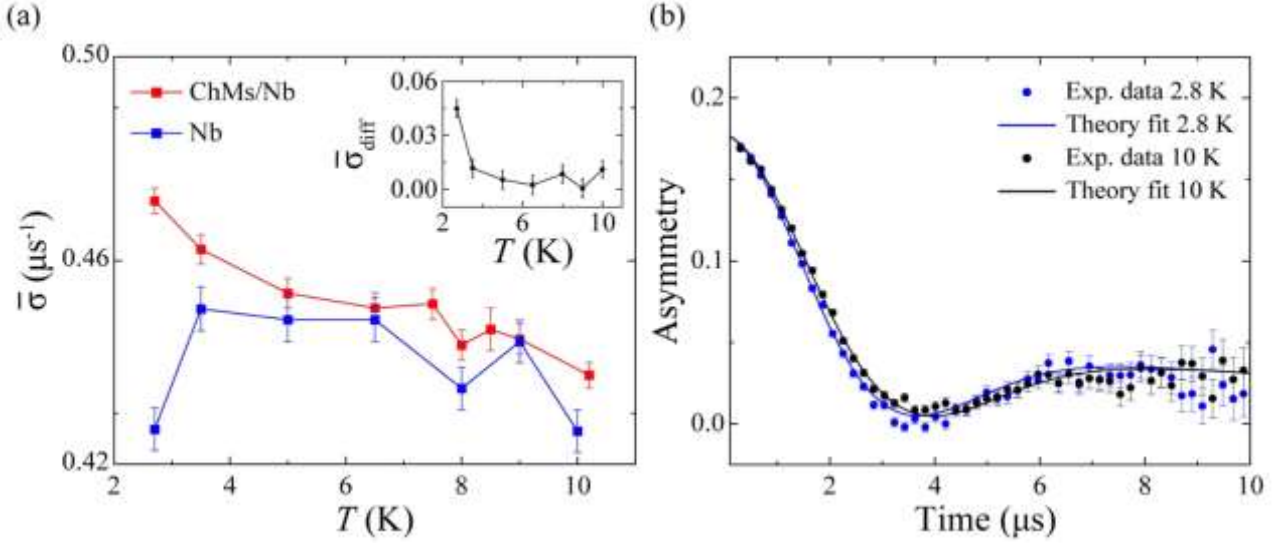

FIG. 10. LE-μSR asymmetry data in zero field. (a) Average muon depolarization rate $\bar{\sigma}$ measured in ZF at $E = 8$ keV across the superconducting transition for a ChMs/Nb (55 nm) sample (red curve) and a bare Nb (55 nm) sample (blue curve). The inset shows the difference in $\bar{\sigma}$ ($\bar{\sigma}_{\text{diff}}$) measured between the two samples as a function of temperature. (b) Asymmetry signal corresponding to the points in (a) measured above $T_c$ at 10 K (black symbols) and below $T_c$ at 2.8 K (blue symbols) with fits to the Kubo-Toyabe model (solid lines with corresponding colors). The variation between the curves is related to an increase in the $\bar{\sigma}$ value of the model, as defined in the main manuscript, and it shows the presence of additional spin activity in the sample with ChMs below $T_c$.

## APPENDIX D: NUMERICAL SIMULATIONS OF THE SCREENING IN CHMs/Nb

As discussed in the main manuscript, the numerical simulation of the screening in the S (Nb) is obtained by solving the Maxwell equations for the vector potential $\boldsymbol{A}$, which takes the form

$$\nabla^2 \boldsymbol{A} = -\mu_0 \boldsymbol{J}, \tag{D1}$$

where $\mu_0$ is the vacuum permeability, $\boldsymbol{J}$ is the electric current density, and the Coulomb gauge has been assumed to be $\nabla \cdot \boldsymbol{A} = 0$. The current density $\boldsymbol{J}$ is determined from the expression

$$\boldsymbol{J} = -\frac{|e|\nu_0 D}{16} \int d\varepsilon \, \text{Tr}[\hat{\rho}_4(\hat{g}^R \nabla \hat{g}^R - \hat{g}^A \nabla \hat{g}^A)] \tanh\frac{\beta\varepsilon}{2}, \tag{D2}$$

where $e$ is the electron charge, $\nu_0$ is the density of states at the Fermi level, $D$ is the diffusion constant, $\varepsilon$ is the quasiparticle energy, $\beta = 1/k_B T$ with $k_B$ the Boltzmann constant and $T$ the temperature, and $\hat{\rho}_4 = \text{diag}(+1, +1, -1, -1)$. The Green's function $\hat{g}^R$ is a $4 \times 4$ matrix with structure in spin and particle-hole space. Furthermore $\hat{g}^A = -\hat{\rho}_4(\hat{g}^R)^\dagger \hat{\rho}_4$.

To find $\hat{g}^R$ we solve the quasiclassical Usadel equation in the superconducting material, which takes the form [38]

$$D\widetilde{\nabla}\cdot\hat{g}^{R}\widetilde{\nabla}\hat{g}^{R}+i[\varepsilon\hat{\rho}_{4}+\widehat{\Delta},\hat{g}^{R}]=0, \tag{D3}$$

where $\widehat{\Delta}=\text{antidiag}(+\Delta,-\Delta,+\Delta,-\Delta)$, $\Delta$ is the superconducting order parameter, and $\widetilde{\nabla}\hat{g}^{R}=\nabla\hat{g}^{R}-i\frac{e}{\hbar}[A\hat{\rho}_{4},\hat{g}^{R}]$. $\Delta$ must be determined through self-consistent iterations using the expression

$$\Delta=-\frac{\nu_{0}\lambda_{k}}{8}\int d\varepsilon\left(\hat{g}_{23}^{R}-\hat{g}_{23}^{A}\right)\tanh\frac{\beta\varepsilon}{2}, \tag{D4}$$

where $\hat{g}_{ij}^{R}$ indicates the element in column $i$ and row $j$ of the matrix $\hat{g}^{R}$, and $\lambda_{k}$ is the superconducting coupling constant. A value of $\lambda_{k}=0.25$ has been assumed.

The adsorbed ChMs are modeled as a spin-active boundary condition, given as [39]

$$\boldsymbol{n}\cdot\hat{g}^{R}\nabla\hat{g}^{R}=-iG_{\phi}[\hat{g}^{R},\boldsymbol{m}\cdot\widehat{\boldsymbol{\sigma}}], \tag{D5}$$

where $\boldsymbol{n}$ is the vector normal to the surface, and $G_{\phi}$ gives the strength of the spin-dependent scattering phase shifts experienced upon reflection at the interface, with spin direction $\boldsymbol{m}$. Furthermore, $\widehat{\boldsymbol{\sigma}}=\text{diag}(\boldsymbol{\sigma},\boldsymbol{\sigma}^{*})$, where $\boldsymbol{\sigma}$ is a vector containing the Pauli matrices. $\boldsymbol{m}$ is assumed collinear with the applied magnetic field $B_{\text{ext}}$.

The numerical model is schematically illustrated in Fig. 11. The width of the S is chosen to be $L=8\xi_{s}$, where $\xi_{s}$ is the superconducting coherence length, and its thickness is $d_{s}=4\xi_{s}$. We assume the presence of a vacuum layer with a thickness of $5\xi_{s}$ surrounding the S material. The spin-active boundary conditions are applied along the top surface of the S indicated by $\Gamma_{\text{so}}$, whereas vacuum boundary conditions, $\boldsymbol{n}\cdot\nabla\hat{g}^{R}=0$, are assumed along the other three surfaces, denoted as $\Gamma_{V}$.

To initialize the solution procedure, the superconducting order parameter is set equal to an initial guess, $\Delta=\Delta_{0}$, and the vector potential is chosen such that it represents the external magnetic field, meaning that $\boldsymbol{A}=\boldsymbol{A}_{0}=\boldsymbol{A}_{\text{ext}}$. Eq. (D3) is then solved in the superconducting material using the finite element method (for details see ref. [40]). This provides the next iteration of $\Delta_{1}$, and the screening current, $J_{1}$, via Eqs. (D2) and (D4), respectively.

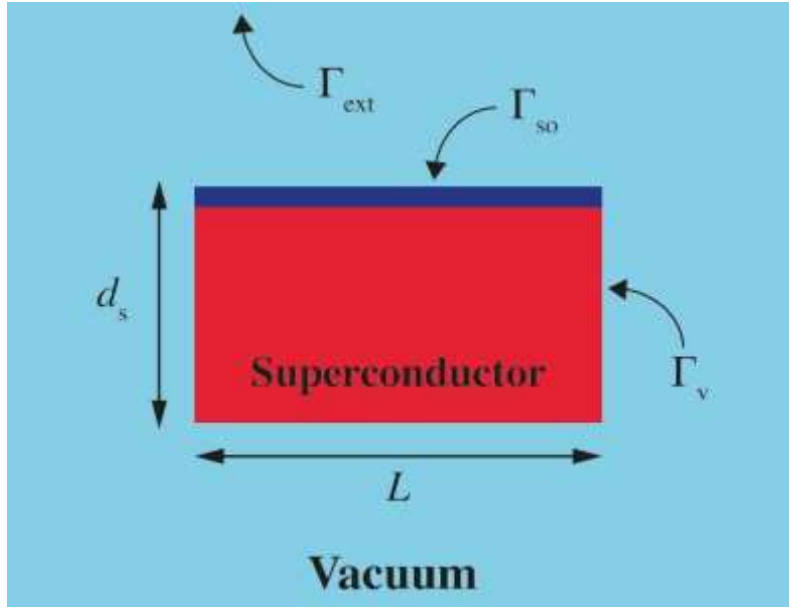

FIG. 11. Theoretical model. Schematic of the theoretical model used to determine the field profile in ChMs/Nb.

Finally, the next iteration of the vector potential, $A_1$, is found by solving Eq. (D1) in both the S and the vacuum regions while at the external boundary of the latter, indicated by $\Gamma_{\text{ext}}$, enforcing the boundary conditions

$$\begin{aligned} \boldsymbol{n} \cdot \nabla(\boldsymbol{n} \cdot \boldsymbol{a}) &= 0, \\ \boldsymbol{a} - \boldsymbol{n}(\boldsymbol{n} \cdot \boldsymbol{a}) &= 0, \end{aligned} \quad (D6)$$

for $\boldsymbol{a} = \boldsymbol{A}_{\text{ext}} - \boldsymbol{A}$. These conditions are equivalent to setting the calculated magnetic field $\boldsymbol{B}$ equal to the applied magnetic field $\boldsymbol{B}_{\text{ext}}$ along $\Gamma_{\text{ext}}$. This procedure is repeated until self-consistency in both OP and the vector potential is achieved. Once both have converged, the magnetization induced by odd-frequency spin-triplet superconducting correlations is calculated from

$$\boldsymbol{M} = \frac{g\mu_B \nu_0}{32} \int d\varepsilon \ \text{Tr}[\hat{\boldsymbol{\sigma}}(\hat{g}^R - \hat{g}^A)] \tanh\frac{\beta\varepsilon}{2}, \quad (D7)$$

where $g \cong 2$ is the Landé $g$ factor, and $\mu_B$ is the Bohr magneton.

The local field profile $B_{\text{loc}}(z)$ that is computed using our theoretical model and normalized to $B_{\text{ext}} = 302.6$ Gauss for a S $d_s = 4\ \xi_s$, it is shown in Fig. 11. It can be inferred from Fig. 11 that $B_{\text{loc}}(z)$, which also corresponds to the yellow curve in Fig. 4(b), properly reproduces the trend of the experimental LE-μSR data reported in Fig. 2(a), whereas a conventional Meissner model (black curve in Fig. 12) cannot properly reproduce the trend followed by the same data set.

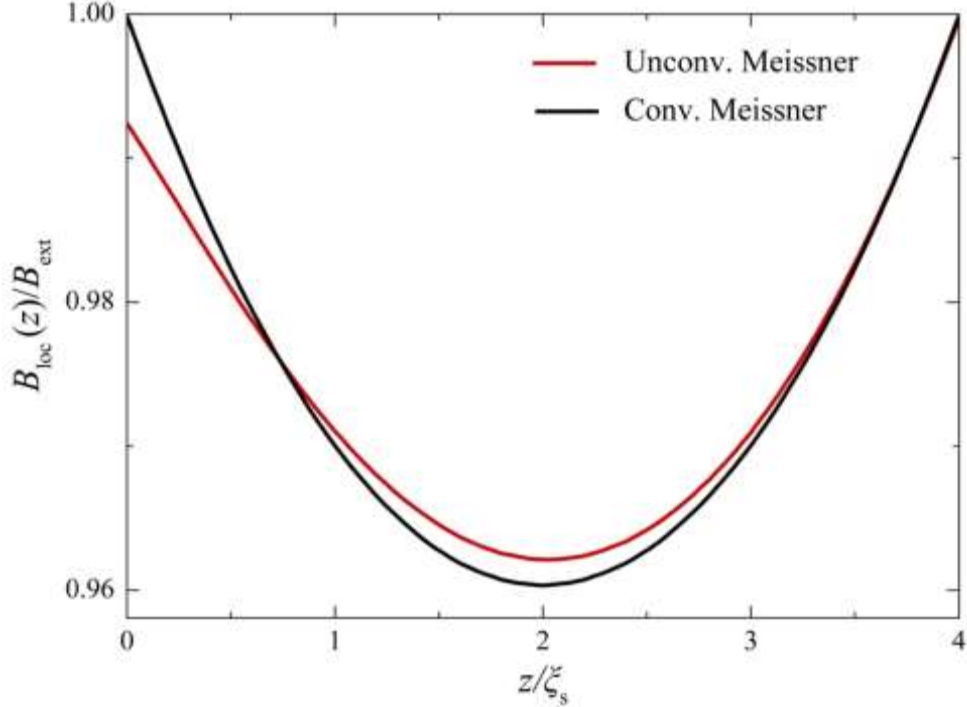

FIG. 12. Theoretical field profile calculated based on our model. Depth dependence of $B_{loc}(z)/B_{ext}$ computed from our theoretical model (red curve) for a ChMs/Nb system with $B_{ext} = 302.6$ Gauss as during the collection of LE-μSR data in Fig. 2. The profile is simulated assuming a Nb thickness $d_s = 4\,\xi_s$. A profile derived from the Ginzburg-Landau model (black curve) for $B_{loc}(z)/B_{ext}$ in the case of conventional Meissner screening is also shown for comparison. We note that $z = 0$ corresponds to the ChMs/S interface in our model, with the ChMs being modeled as the equivalent of a spin-active insulator. At the interface between the surface of the spin-active insulator and vacuum (not shown here), the local field in our model coincides with $B_{ext}$.

## APPENDIX E: FITTING PROCEDURE FOR THE LE-μSR DATA

### 1. Theory and single-energy fitting of the LE-μSR data

A schematic illustration of the experimental apparatus used to probe the local magnetic field profile $B_{loc}(z)$ in the ChMs/Nb and bare Nb samples (with Nb thicknesses of 65 nm and 55 nm) is given in Fig. 1. In this section, we describe how the analysis is carried out for LE-μSR measurements done in a transverse field (TF) configuration, where the initial muon polarization is perpendicular to the applied external field $B_{ext}$.

The starting point for the analysis of the muon data is the asymmetry signal $A_s(t, E)$ which is experimentally determined from number of events $N(t, E)$ counted by the left and right arrays of positron detectors (in the following, we denote the parameters referring to the left and right detectors with the subscripts 'L' and 'R', respectively). The number of positron events $N$ (or equivalently of muon decaying events) is related to $A_s(t, E)$ by the expressions

$$N_L(t, E) = N_0 e^{-t/\tau_\mu}[1 + A_s(t, E)] + N_{bkg,L}, \tag{E1}$$

and

$$N_R(t,E) = \alpha_d N_0 e^{-t/\tau_\mu}[1 - A_s(t,E)] + N_{bkg,R}, \tag{E2}$$

where $N_{bkg}$ is the time-independent background contribution due to accidental coincidences, $\tau_\mu \sim 2.2$ μs is the muon lifetime, and $\alpha_d \sim 1$ is a correction factor for detector efficiency.

The signal $A_s(t,E)$ can be determined from Eqs. (E1) and (E2) because it corresponds to the difference of the counting events between left and right detectors divided by their sum meaning that

$$A_s(t,E) = \frac{\alpha_d[N_L(t,E) - N_{bkg,L}] - [N_R(t,E) - N_{bkg,R}]}{\alpha_d[N_L(t,E) - N_{bkg,L}] + [N_R(t,E) - N_{bkg,R}]}. \tag{E3}$$

The asymmetry function is proportional to the muon polarization. In particular, $A_s(t,E)$ can be written in its simplest form as

$$A_s(t,E) = A_{s0} \cos\left(\gamma_\mu \bar{B}_{loc}(\bar{z}(E))t + \varphi_0(E)\right) G(t,E), \tag{E4}$$

with $\gamma_\mu = 2\pi \cdot 135.5$ MHz Tesla$^{-1}$ being the muon gyromagnetic ratio, $A_{s0}$ the initial asymmetry amplitude, $\varphi_0(E)$ the initial phase of the muon precession, and $G(t,E)$ the depolarization function due to inhomogeneities and/or dynamics in the local field experienced by muons at their implantation sites. Eq. (E4) includes the broadening effect of the muon stopping distribution $p(z,E)$ on the determination of $\bar{B}_{loc}(\bar{z}(E))$, since $\bar{B}_{loc}(\bar{z}(E))$ is the weighted average of the actual local field $B_{loc}(z)$ over $p(z,E)$.

Eqs. (E3) and (E4) are used in combination to perform a single-energy asymmetry fit at a specific energy $E$. For all the ChMs/Nb and Nb samples, the best fits were obtained with a Gaussian $G(t,E)$ function meaning with $G(t,E) = e^{-1/2(\bar{\sigma}^2/t^2)}$, where $\bar{\sigma}$ is the muon depolarization rate.

In Fig. 13 below, we show some representative single-energy asymmetry fits based on the model described above. The representative asymmetry fits are shown for both ChMs/Nb (65 nm) and Nb (65 nm) at two different energies ($E$ = 3 keV, 7.5 keV) and two different temperatures ($T$s), of which one below $T_c$ ($T$ = 3 K) and the other one above $T_c$ ($T$ = 10 K). Fig. 13 shows that the fits accurately reproduce the raw data, as also evidenced by the very low chi-square/number of degrees of freedom (**chisq/NDFs**) of ~ 1.02 obtained in the fitting software musrfit. By performing single-energy asymmetry fits like those shown in Fig. 13 for a few representative cases, we determine the sequence of $\bar{B}_{loc}(\bar{z}(E))$ values as a function of $E$ such as that shown, for example, in Fig. 2(a), which provides a preliminary estimate of the local magnetic field profile $B_{loc}(z)$ inside the sample.

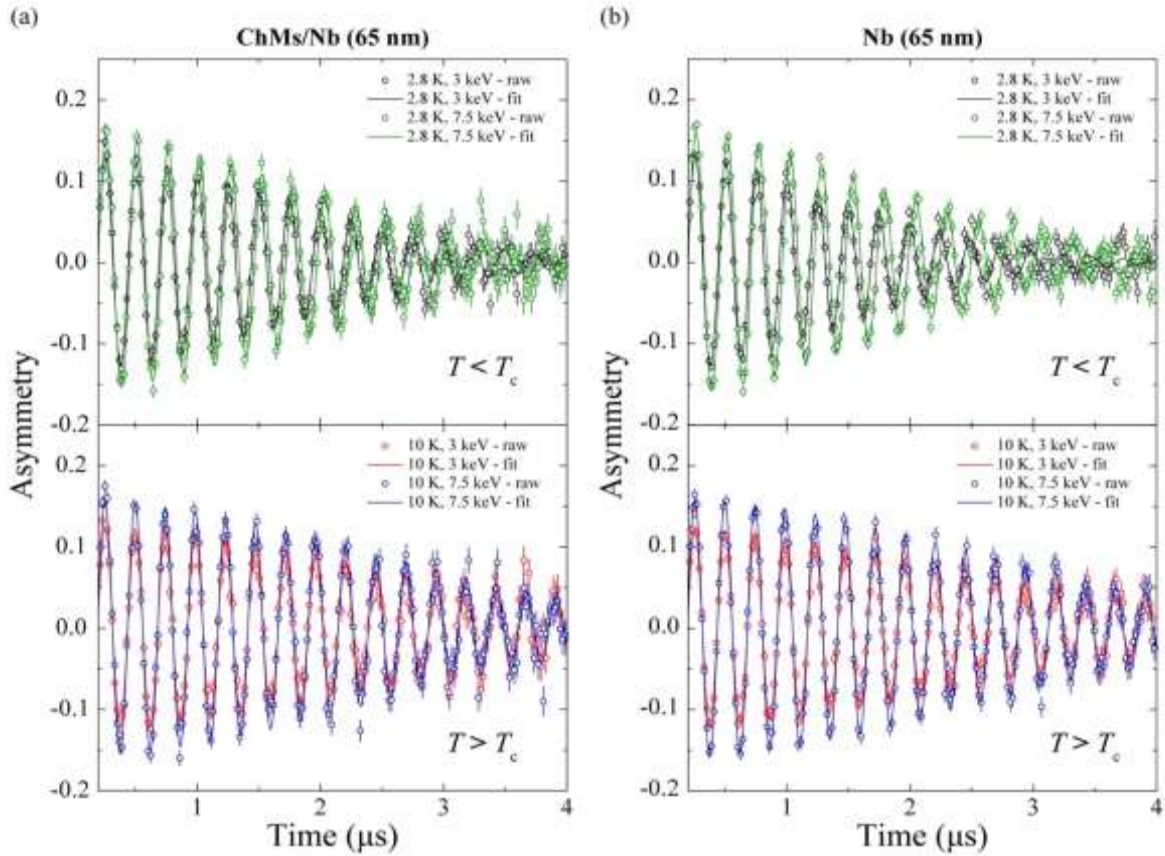

FIG. 13. Asymmetry signal and corresponding fits. (a)-(b), Examples of raw data (symbols with error bars) and theoretical fits (solid lines) for the asymmetry signal $A_S(t)$ measured for a ChMs/Nb (65 nm) sample (a) and Nb (65 nm) sample (b). The top panels in (a) and (b) show the raw $A_S(t)$ data and corresponding fits obtained for both samples below $T_c$ at $T = 2.8$ K and $E = 3$ keV (black symbols and lines) or $E = 7.5$ keV (green symbols and lines). The bottom panels in (a) and (b) show the raw $A_S(t)$ data and corresponding fits obtained for both samples above $T_c$ at $T = 10$ K and $E = 3$ keV (red symbols and lines) or $E = 7.5$ keV (blue symbols and lines).

We note that we also perform transverse-field LE-μSR measurements in $B_{ext} = 300$ Gauss on the same ChMs/Nb (55 nm) sample on which the zero-field LE-μSR measurement data shown in Fig. 4 are collected. The $\bar{B}_{loc}(\bar{z})$ profiles for this sample, which are also determined as sequence of the single-energy asymmetry fits as a function of $E$, are reported in Fig. 14. The results in Fig. 14 show that also for ChMs/Nb (55 nm) in the superconducting state at 2.8 K we can reproduce the same enhancement in Meissner screening at the ChMs/Nb interface measured for the two different ChMs/Nb (65 nm) samples and reported in Fig. 2 and Fig. 3.

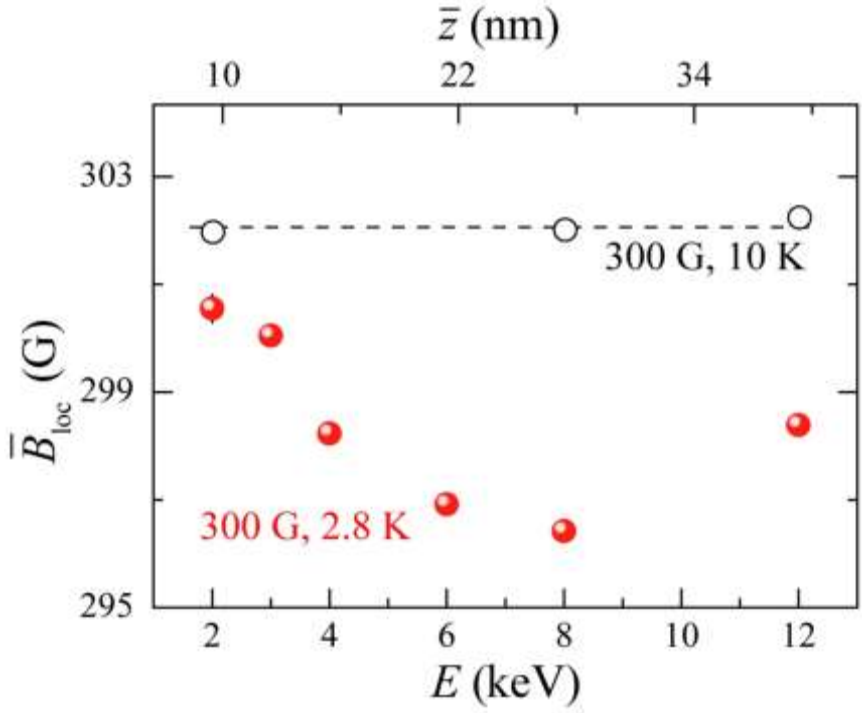

FIG. 14. Magnetic field response in ChMs/Nb (55 nm). Average local magnetic field $\bar{B}_{loc}$ as a function of muon implantation energy $E$ (bottom axis) and average muon stopping depth $\bar{z}$ (top axis) determined by transverse-field LE- μSR measurements in the superconducting state ($T_c \sim 8.7$ K) at $T = 2.8$ K (red symbols) and in the normal state at $T = 10$ K (black hollow symbols).
The $\bar{B}_{loc}$ profile at $T = 2.8$ K shows an unconventional Meissner screening consistent with that measured for the two other ChMs/Nb (65 nm) samples shown in Fig. 2(a) and Fig. 3.

## 2. Global fitting and analytical model

To obtain a more accurate $B_{loc}(z)$, a common fit for all energy runs that takes into account also the contribution of the muon stopping profiles $p(z, E)$ is performed. For the global fit, Eq. (E4) is replaced by the expression

$$A_s(t, E) = \int p(z, E) A_{s0} \cos\left(\gamma_\mu B_{loc}(z) t + \varphi_0(E)\right) G(t, E) \, dz, \tag{E5}$$

where the integral is extended to the whole depth range probed by muons at a given implantation energy $E$. In Eq. (E5), $B_{loc}(z)$ is not treated as a constant value (as for Eq. (E4)) but it is given by an analytical expression that is also fed as input to the fitting algorithm (along with the simulated $p(z, E)$ which are shown for several energies in Fig. 15). The aim of the global fit is therefore to find an analytical model for $B_{loc}(z)$ which is physically meaningful and that, at the same time, properly fits the experimental data measured for all $E$s.

For the bare Nb (65 nm) sample (i.e., without the ChMs), in the global fit we use as analytical model for $B_{loc}(z)$, which we define as $B_{loc, conv.}(z)$, the expression given by the Ginzburg-Landau phenomenological theory of superconductivity meaning

$$B_{loc, conv.}(z) = B_{ext} \cosh\left(\frac{z - \frac{d_s}{2}}{\lambda_{GL}}\right) / \cosh\left(\frac{d_s}{2\lambda_{GL}}\right), \tag{E6}$$

where $\lambda_{GL}$ is the Ginzburg-Landau penetration depth and $d_s$ the thickness of the S layer.

For the global fit instead for the ChMs/Nb (65 nm) system, we model the local field with an analytical expression $B_{\text{loc, unc.}}(z)$ that qualitatively captures the main physical picture described by our theoretical model for the unconventional screening in ChMs/Nb. The analytical expression for $B_{\text{loc}}(z)$ which we adopt has the form

$$B_{\text{loc, unc.}}(z) = B_{\text{Meissner}}(z) + B_{\text{spin}}(z), \tag{E7}$$

where $B_{\text{Meissner}}(z)$ is a modified version of $B_{\text{loc, conv.}}(z)$ for conventional Meissner screening given by Eq. (E6). In particular, as also explained in the main manuscript, in $B_{\text{Meissner}}(z)$ we introduce a spatial dependence for $\lambda(z)$ to account for the variation in the screening properties of the S whilst moving away from the interface with ChMs – which we model theoretically as a spin-active insulating layer. The two terms in Eq. (E7) are therefore explicitly defined as

$$B_{\text{Meissner}}(z) = B_{\text{ext}} \frac{\cosh\left(\frac{z-\frac{d}{2}}{\lambda(z)}\right)}{\cosh\left(\frac{d}{2\lambda(z)}\right)} \tag{E8}$$

with

$$\lambda(z) = \frac{\lambda_{\text{GL}}}{1+\alpha\, e^{(-z/\xi_{\text{lambda}})}} \tag{E9}$$

and as

$$B_{\text{spin}}(z) = B_{\text{spin}} e^{(-z/\xi_{\text{spin}})}. \tag{E10}$$

In Eq. (E9), $\xi_{\text{lambda}}$ is the length scale over which $\lambda(z)$ restores to $\lambda_{\text{GL}}$, meaning to the penetration depth value that would be measured in Nb without ChMs. $\xi_{\text{spin}}$ in Eq. (E10) defines instead the length scale over which the $B_{\text{spin}}(z)$ decays from the ChMs/Nb interface. Using Eqs. (E9) and (E10), the expression given by Eq. (E7) can be therefore explicitly written as

$$B_{\text{loc, unc.}}(z) = B_{\text{ext}} \frac{\cosh\left(\frac{z-\frac{d}{2}}{\lambda(z)}\right)}{\cosh\left(\frac{d}{2\lambda(z)}\right)} + B_{\text{spin}} e^{(-z/\xi_{\text{spin}})}. \tag{E11}$$

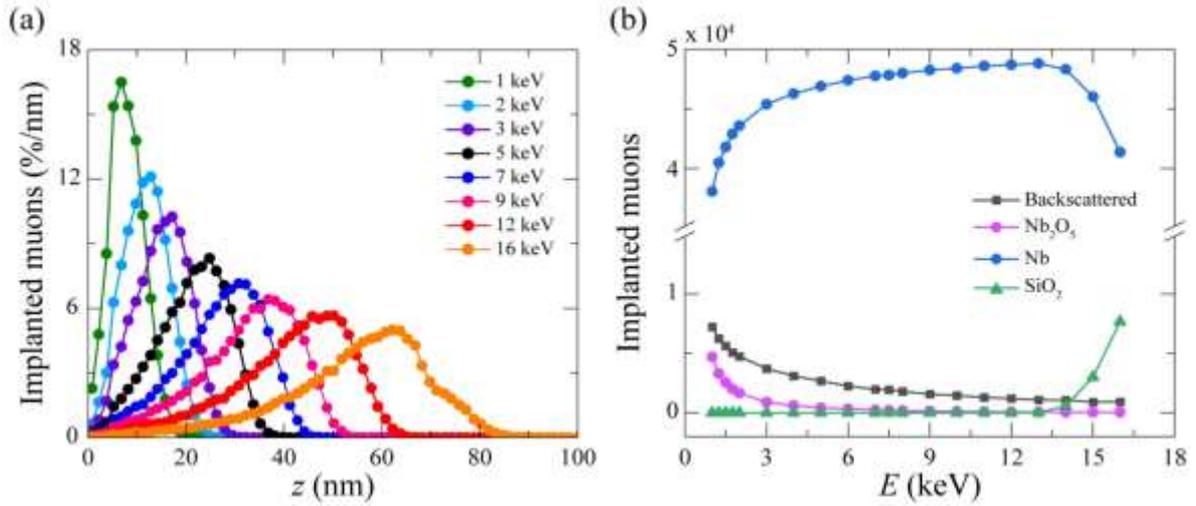

FIG. 15. Simulated normalized stopping profiles and fraction of implanted muons at different energies. (a)-(b), Normalized muon stopping profile distributions $p(z, E)$ for Nb (65 nm) thin film with a top 4-nm-thick $Nb_2O_5$ native oxide layer simulated for different implantation energy $E$ values (a) and fraction of implanted muons stopping in $Nb_2O_5$ (magenta curve), Nb (blue curve), in the $SiO_2$ substrate (green curve) and of backscattered muons (grey curve) (b). The thicknesses for the $Nb_2O_5$ and Nb layers are obtained from X-ray reflectometry (Fig. 9) done on the same samples used for the LE-µSR measurements. The layer of ChMs is not included in the simulations due its negligible (monolayer) thickness and low stopping power (due to its low density) for the implanted muons.

*Consideration about the boundaries of the range for some parameters used*

The parameter $\alpha$ in Eq. (E9) should vary between -1 and 0 since $\lambda(z)$ at the ChMs/Nb interface (i.e., at $z = 0$) must larger than $\lambda_{GL}$ according to our theoretical model – this is equivalent to assuming a suppression of $\Delta$ at the interface with ChMs due to their magnetic activity.

$B_{spin}(z)$ in Eq. (E10) represents the magnetization induced in the S by odd-frequency spin-triplet correlations generated at the ChMs/S interface. Our theoretical simulations, consistently with the $\bar{B}_{loc}(\bar{z})$ profile from single-energy asymmetry fits in Fig. 2(a), suggest that the amplitude of $B_{spin}$ at $z = 0$ in Eq. (E10) must be negative.

*Consideration about fitting parameters and fitting procedure*

To perform a global fit to the experimental LE-µSR data, we use the software musrfit [27]. In the input (.msr) file to the fitting routine implemented by musrfit, we define some parameters that are common to all the energy ($E$) runs, and some other parameters that are instead $E$-specific. The parameters that are common to all energies are named in the input file as:

- $z_{start} = 0$ which corresponds to the coordinate of the top surface of the sample.
- $z_{dead}$ which corresponds to the thickness (in nanometers) of the oxide passivation layer forming on the Nb surface plus any other additional layer where the distribution of screening supercurrents is not homogeneous. In other terms, $z_{dead}$ defines the thickness of a layer from the top sample surface, where the field drop from the external field is slower than the exponential one expected according to Ginzburg-Landau theory (see below for further details about this parameter).

- $z_{end}$ that represents the coordinate of the bottom interface of the superconducting portion of the Nb sample.
- $z_{dead2}$ which corresponds to the coordinate of the bottom dead layer at the interface with the SiO$_2$ substrate (i.e. $z_{dead2}$-$z_{end}$ is the thickness of the bottom dead layer).
- *B_ext* corresponding to $B_{ext}$ in Eq. (E8) which is the applied field.
- *B_spin* which corresponds to the $B_{spin}$ amplitude in Eq. (E10).
- *lambda_GL* which represents $\lambda_{GL}$ in Eq. (E9).
- *amp* which corresponds to the parameter $\alpha$ in Eq. (E9).
- *xi_lambda* and *xi_spin* which are $\xi_{lambda}$ and $\xi_{spin}$ in Eqs. (E9) and (E10), respectively.
- *rate_Nb* which is related to the depolarization of muons due to Nb nuclear dipole moments.
*phase_L* and *phase_R* which represent the initial phase of the muons with respect to the left and right arrays of detectors (see Fig. 1), respectively. These parameters are determined by the measurement setup.

Some of the above-listed parameters are *kept fixed* in the fitting routine. One of these parameters is *rate_Nb* which is a material-dependent parameter, since it is related to the depolarization of the muons induced by Nb nuclear dipole moments. The value of such parameter can be estimated from the single-energy asymmetry fits the normal-state LE-µSR data acquired at $T$ = 10 K for bare Nb, for which we find *rate_Nb* ~ 0.41. From a global fit including all the data points at different $E$s in the normal state for the bare Nb (65 nm) sample we verify, however, that a slightly larger value of *rate_Nb* ~ 0.45 results in a better fit, meaning in a fit with a lower chisq/NDF going from 1.11 down to 1.07 as *rate_Nb* is increased from 0.41 to 0.45. It is reasonable to expect a slight increase in *rate_Nb* to account for imperfections in the model due, for examples, to inaccuracies in the simulated muon stopping distributions. Based on these considerations, we fix *rate_Nb* = 0.45 for the global fit of the data collected in the superconducting state on ChMs/Nb.

For the boundaries of the Nb film, we set $z_{start}$ = 0, since the Nb surface corresponds to the origin of the $z$-axis according to the reference system adopted in Fig. 1. We are also aware that a Nb$_2$O$_5$ oxide layer naturally forms by passivation when the Nb surface is exposed to air [41]. In this oxide layer, which we determine to be ~ 4 nm in thickness from the X-ray reflectometry data in Fig. 9, no superconducting screening currents should be present and $B_{loc}(z) = B_{ext}$. As reported in previous LE-µSR experiments [20,42], roughness at the interface between the oxide layer and the S underneath results in a layer, also called "dead layer", where the screening current distribution sensed by muons is disordered. We model the dead layer via the parameter $z_{dead}$, which we expect to be larger than the thickness of the top Nb$_2$O$_5$ layer due to roughness at the Nb$_2$O$_5$/Nb interface, consistently with what reported in refs. [20,42]. Inside the dead layer, the decay in $B_{loc}(z)$ (from $B_{ext}$) is slowed due to the inhomogeneous screening currents flowing herein and it does not follow a proper exponential drop from the S surface. Therefore, for $z_{start} < z < z_{dead}$, we assume that $B_{loc}(z) \sim B_{ext}$.

The considerations made for the dead layer at the (top) Nb surface also apply to the Nb/SiO$_2$ interface, where a thin Nb$_2$O$_5$ layer (~ 2 nm in thickness) also forms during deposition of Nb on the oxide substrate [37]. We assume the presence of this second dead layer at the Nb/SiO$_2$ interface by setting a $z_{end}$ value smaller than the nominal thickness of Nb – which we know to be 65 nm from the XRR measurements in Fig. 9. We note that this second dead layer takes into account also any contributions to the measured asymmetry coming from muons stopping in the SiO$_2$ substrate.

$$B_{loc, unc.}(z) = \begin{cases} B_{ext}, & 0 < z < z_{dead} \\ B_{Meissner}(z) + B_{spin}(z), & z_{dead} < z < z_{end} \\ B_{ext}, & z_{end} < z < z_{dead2} \end{cases} \quad (E12)$$

**3. Global fit for the bare Nb sample**

We use the same format for the musrfit input file both for the global fit for the ChMs/Nb sample and for the global fit for the Nb sample. We perform the fit through a multi-step approach. In all the steps, *B_ext* = 302.6 Gauss is fixed (see above) and we use *phase_L* and *phase_R* as global parameters (to fit) common to all energy runs. For the other variable parameters to fit, which are $z_{dead}$, $z_{end}$, $z_{dead2}$ and $\lambda_{GL}$, in each step we do not vary more than three parameters at a time.

At the end of the last step of the fitting routine, we get convergence with a chi square/number of degrees of freedom (chisq/NDF) = 1.117 with the following parameter values: $z_{dead}$ =12.22 ± 0.2 nm, $z_{end}$ = 55.30 ± 0.88 nm, $z_{dead2}$ = 62.1 nm (fixed) and $\lambda_{GL}$ = 66.4 ± 1.2 nm.

The curve obtained with these parameter values convoluted with the muon stopping distributions yields a good representation of the $\bar{B}_{loc}(\bar{z})$ profile from the single-energy asymmetry fit (see Fig. 5(a)), which validates the outcome of the global fit algorithm for the bare Nb sample.

**4. Global fit for the ChMs/Nb**

Similar to the case of bare Nb (65 nm), also for the global fit of the LE-μSR data on ChMs/Nb (65 nm) we follow a multistep approach, where we minimize the number of fitting parameters used in each step. In each step, as for the bare Nb case, we keep *B_ext* = 302.6 Gauss fixed and use *phase_L* and *phase_R* as global parameters (to fit) common to all energy runs.

In the first step of the fitting routine, we set $z_{dead}$, $z_{end}$, $z_{dead2}$ fixed to the values obtained from the global fit on bare Nb (65 nm). The assumption to use these values as starting point for the fit is reasonable because the two Nb films are twin samples deposited simultaneously in the same conditions. In addition, also the bare Nb (65 nm) sample was dipped overnight in an ethanol bath (but without ChMs) before loading it into the LE-μSR setup for measurements. This was done in order to follow the same exact sample preparation steps, not only during growth, but also before measurements, for both the ChMs/Nb (65 nm) and Nb (65 nm) samples, in order to rule out that possible differences in the LE-μSR measurements between the two samples would be caused by different experimental conditions adopted for them. The immersion of a Nb film in an ethanol bath, for example, may also contribute to increase the thickness of the $Nb_2O_5$ oxide layer as reported in ref. [44]. For the other variable parameters ($\lambda_{GL}$, $B_{spin}$ and *xi_spin*) we get convergence, after the first step, with the values: $\lambda_{GL}$ = 74.72 $\pm$ 0.59 nm, $B_{spin}$ = -6.00 $\pm$ 0.34 Gauss and *xi_spin* = 7.7 $\pm$ 0.81 nm.

In the second step of the fitting routine, we fix *xi_spin*, $B_{spin}$ and $\lambda_{GL}$ to the values found in the previous step, and we do a fine tuning of the parameters $z_{dead}$ and $z_{end}$ which yields $z_{dead}$ = 11.81 $\pm$ 0.18 nm and $z_{end}$ = 55.44 $\pm$ 0.18 nm – which are very close to the values obtained from the global fit for bare Nb, as one would expect since the Nb thin films were grown in the same deposition run. This does not only suggest that the thickness of the dead layer is reproducible between samples grown in the same run, but it also suggests that the variations that we observe in the local field profiles of ChMs/Nb (65 nm) and bare Nb (65 nm) shown in Fig. 2 cannot be related to any effects due to the dead layer. We also note that the results in Fig. 3 showing the asymmetry of the unconventional Meissner screening in ChMs/Nb (65 nm) under switching of the direction of $B_{ext}$ also rule out that this unconventional Meissner response is originated by effects due to the dead layer.

In the last step of the fitting routine, we keep $z_{dead}$ = 11.81 nm (fixed), $z_{end}$ = 55.44 nm (fixed), $z_{dead2}$ = 62.1 nm (fixed) and use $\lambda_{GL}$, $B_{spin}$ and *xi_spin* as fitting parameters. We obtain convergence with chisq/NDFs = 1.093 and the following parameter values: $\lambda_{GL}$ = 73.9 $\pm$ 1.0 nm, $B_{spin}$ = -6.6 $\pm$ 1.1 Gauss and *xi_spin* = 6.0 $\pm$ 2.0 nm. We also verify, that the correction to $\lambda_{GL}$ introduced in Eq. (E9) through the term $\alpha\, e^{(-z/\xi_{lambda})}$ is negligible for $z_{dead} < z < z_{end}$. The fit returns a value of $\alpha$ very close to zero, meaning that this term can be neglected beyond $z_{dead} \sim$ 10-11 nm from the Nb surface, in agreement with the results of our theoretical simulations.

We note that the $\lambda_{GL}$ value obtained from the fit for ChMs/Nb (65 nm) is slightly larger than the value obtained from the fit for Nb (65 nm). This small difference is due to a combination of small sample-to-sample variations in $\lambda_{GL}$ and of the approximation that we make that that λ(z) is constant and depth independent for the ChMs/Nb sample. As a result of this approximation, the fit should return a constant value in between the value which λ(z) takes at the ChMs/Nb interface, which must be larger than $\lambda_{GL}$ for bare Nb due to the gap suppression at the ChMs/Nb interface, and the value that λ(z) value takes deep inside Nb, where λ(z) should recover to $\lambda_{GL}$ for bare Nb.

After substituting the parameter values listed above in Eq. (E11) and convoluting the as-obtained curve by the muon stopping profiles in Fig. 14(a), we get the red solid curve in Fig. 5(a) of the main manuscript – which represents a good fit to the single-energy asymmetry fit data acquired on the ChMs/Nb (65 nm) and therefore demonstrates the validity of our theoretical model in describing the unconventional Meissner screening in ChMs/Nb as well as the validity of the fitting procedure followed.

## 5. Analysis of the zero-field LE-μSR measurements

The dynamic depolarization rate as a function of temperature for both the ChMs/Nb (55 nm) and bare Nb (55 nm) is presented in Fig. 4 and Fig. 10. In the absence of external magnetic field, muons will precess along the local field generated by the magnetic moments of the Nb nuclei and other static magnetic moments. Due to the polycrystalline nature of the Nb films, the dense (and randomly oriented) nuclei moments average out to generate a normal field distribution of a Gaussian centered about zero field: $f(B_j) = \frac{1}{\sqrt{2\pi\langle\Delta B^2\rangle}} exp\left[-\frac{B_j^2}{2\langle\Delta B^2\rangle}\right]$, where $j = x, y, z$ are the cartesian coordinates. By substituting this normal distribution into the muon depolarization equation, $P_z(t) = \int f(\vec{B})[\cos^2(\theta) + \sin^2(\theta)\cos(\gamma_\mu B_\mu t)]d\vec{B}$, one gets the Kubo-Toyabe depolarization function [32] which has the form

$$P_z(t) = \frac{1}{3} + \frac{2}{3}(1 - \bar{\sigma}^2 t^2)\exp\left[-\frac{\bar{\sigma}^2 t^2}{2}\right]. \tag{E13}$$

The $P_z(t)$ depolarization function above is therefore related to the magnetic dipolar fields generated by the nuclei and other static moments present inside the sample. In the Kubo-Toyabe model, the $P_z(t)$ function is multiplied by a decaying exponential term: $e^{-\bar{\nu}t}$. The exponential term $e^{-\bar{\nu}t}$ accounts for stochastic dynamical effects with field autocorrelation time of $\frac{1}{\bar{\nu}}$, meaning that, after a certain time $t = \frac{1}{\bar{\nu}}$, a fluctuating moment assumes a random value taken from a Gaussian field distribution $f(B)$ centered about zero field. For a static field with zero fluctuations the, Kubo-Toyabe function is recovered, and in the limit that $\bar{\nu} \ll \bar{\sigma}$ the muon polarization becomes

$$P_z(t) = e^{-\bar{\nu}t}\left(\frac{1}{3} + \frac{2}{3}(1 - \bar{\sigma}^2 t^2)\exp\left[-\frac{\bar{\sigma}^2 t^2}{2}\right]\right), \tag{E14}$$

Combining Eqs. (E3) and (E14), we therefore fit the experimental asymmetry measured in ZF using the expression

$$A_s(t, E) = A_{s0} e^{-\bar{\nu}t}\left(\frac{1}{3} + \frac{2}{3}(1 - \bar{\sigma}^2 t^2)\exp\left[-\frac{\bar{\sigma}^2 t^2}{2}\right]\right), \tag{E15}$$